\documentclass[sigconf]{acmart}

\usepackage{booktabs} % For formal tables

\usepackage{amssymb}
\usepackage{graphicx}

\usepackage{url}

\usepackage{comment}
\usepackage{epsfig}
\usepackage{multirow}
\usepackage{booktabs}
\usepackage{diagbox}
\usepackage{bbding}
\usepackage{textcomp,booktabs}

\usepackage{bmpsize}

\usepackage{subfigure}
\usepackage{epstopdf}
\usepackage{algorithm}
\usepackage[noend]{algpseudocode}

\makeatletter
\def\BState{\State\hskip-\ALG@thistlm}
\makeatother

\usepackage{pifont}% http://ctan.org/pkg/pifont
\usepackage{tikz}

\usepackage{hyperref}
\usepackage{epstopdf}
\usepackage{algorithm}
\usepackage[noend]{algpseudocode}

\makeatletter
\def\BState{\State\hskip-\ALG@thistlm}
\makeatother

\usepackage{pifont}% http://ctan.org/pkg/pifont
\usepackage{array}
\usepackage{amsmath}
\usepackage{bm}
\acmArticle{4}
\begin{document}

%\title{Seq2seq Translation Modeling \\ for Context-Aware Sequential Recommendation}
\title{Seq2seq Translation Model for Sequential Recommendation}
%\begin{comment}
\author{Ke Sun}
\affiliation{%
\institution{Wuhan University}
\streetaddress{}
\city{Wuhan}
\state{Hubei}
\country{China}
}
\email{sunke1995@whu.edu.cn}

\author{Tieyun Qian}
\authornote{Corresponding author}
\affiliation{%
\institution{Wuhan University}
\streetaddress{}
\city{Wuhan}
\state{Hubei}
\country{China}
}
\email{qty@whu.edu.cn}
%\end{comment}

\begin{abstract}
The context information such as product category plays a critical role in sequential recommendation. Recent years have witnessed a growing interest in context-aware sequential recommender systems. Existing studies often treat the contexts as auxiliary feature vectors without considering the sequential dependency in contexts. However, such a dependency provides valuable clues to predict the user's future behavior. For example, a user might buy electronic accessories after he/she buy an electronic product.

In this paper, we propose a novel seq2seq translation architecture to highlight the importance of  sequential dependency in contexts for sequential recommendation. Specifically, we first construct a collateral context sequence in addition to the main interaction sequence. We then generalize recent advancements in translation model from sequences of words in two languages to sequences of items and contexts in recommender systems. Taking the category information as an item's context, we develop a basic coupled and an extended tripled seq2seq translation models to encode the category-item and item-category-item relations between the item and context sequences. We conduct extensive experiments on two real world datasets. The results demonstrate the superior performance of the proposed model compared with the state-of-the-art baselines.
\end{abstract}

%\ccsdesc[500]{Computer systems organization~Embedded systems}


\maketitle

\section{Introduction}
Recommendation system has become an inseparable part of our daily lives in the era of information explosion. A good recommendation system works like an information filter which can learn users' interests based on their profile or preferences and then make proper predictions on their future behaviors. There are two main types of conventional recommendation systems, i.e., general recommenders and sequential recommenders.
Researches on general recommendation  mainly focus on modeling users' general static preferences. Methods like matrix factorization \cite{koren2009matrix} and BPR \cite{rendle2009bpr} have shown impressive success towards this direction. However, in real-world scenarios, users' future behavior can be greatly influenced by their current actions. Hence more recent studies have paid considerable interests on modeling such sequential behaviors.

The core of sequential recommendation lies in capturing the latent transition dependencies in users' historical records. Many approaches \cite{tang2018personalized,yu2016dynamic,wang2015learning,kang2018self} try to model not only the interactions between users and items but also the evolution of users' dynamic interests, and can get more accurate predictions compared with traditional methods. Markov Chains  (MCs) are initially employed by building a transition matrix for sequential recommendation  \cite{rendle2010factorizing}. Afterwards, with the prevalence of deep learning techniques, many deep neural networks such as recurrent neural networks (RNN), convolutional neural network (CNN), and attention mechanism have also been incorporated  into sequential recommender systems \cite{yu2016dynamic,tang2018personalized,kang2018self}. However, all these methods only concentrate on the item sequence itself without considering the rich context information.

The context information can provide new perspectives to understand users' intrinsic intentions, and it has been proved helpful in improving the performance of sequential recommendation. Contexts can be viewed from both users' and items' perspective.
The user's contexts mainly consist of  user's profile information like age or profession and user's actions like clicks ~\cite{yao2017serm,zhou2018atrank,le2018modeling}. It is usually hard to access users' contexts due to the privacy protection issues, thus researchers pay more attention to items' contexts ~\cite{he2017category,bai2018attribute,huang2018csan,huang2018improving,chang2018content} such as category, brand, image, descriptions, or the location of a venue.

Existing studies have made considerable progress in modeling context information in sequential recommendation. However,  most of these studies aim at learning fine-grained user or item representations with the help of contexts. For example, a number of approaches extract contextual features before sending them to the sequential recommendation module ~\cite{yao2017serm,bai2018attribute,zhou2018atrank,chen2019air}, and some approaches employ context-specific matrices to capture the temporal and spatial contexts in location prediction ~\cite{liu2016context,liu2016predicting}.
The aforementioned methods are all built upon the single item sequence without considering the latent transition patterns in contexts. In other words, they concentrate on the sequential modeling of the items and ignore the dynamics of additional contexts. It is worth mentioning that
though STAR ~\cite{rakkappan2019context} apples two RNN sequences or two matrices to model the sequences of item and context respectively, the transition patterns in contexts are not explicitly exploited. Moreover, the context and item sequence are treated separately  without considering their relations in these two sequences.
\begin{comment}
Some make use of deep memory network to incorporate the category\cite{huang2019taxonomy,chen2019air} or knowledge base information\cite{huang2018improving}. Some others directly adapt the RNN architecture to the temporal context with deliberately designed transition matrices\cite{liu2016context,liu2016predicting}.

Being aware of the context information's significance, we can divide it into three different types: user profile (e.g., age, profession), user's evaluation on the item (e.g., review, rating) and item attribute (e.g., category, brand, image). According to the current literature, existing approaches\cite{huang2018csan,chang2018content,yao2017serm,bai2018attribute,zhou2018atrank,huang2018improving,chen2019air,huang2019taxonomy} are major in modeling the last two kinds of context information  and have made non-negligible progress on the next-item prediction problem.
\end{comment}

\begin{figure}[htb]
	\centering
	\vspace{-0.3cm}
	\hspace{-0.35cm}\includegraphics[width=0.38\textwidth,height=4.2cm]{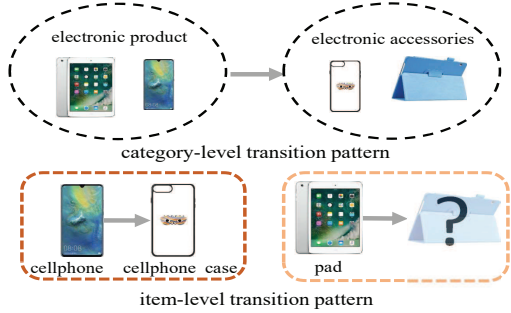}
	\vspace{-0.1cm}
	\caption{The category- and item- level transition patterns.}\label{fig:example}
	\vspace{-0.3cm}
\end{figure}

In this paper, we argue that the sequential dependency in contexts provide valuable clues to predict the user's future intentions. We first present an example of category level and item level transition pattern in Figure ~\ref{fig:example}. We choose the category as an items' context since the category information is easily accessible in most of real world e-commerce web sites or online social networks. Also, the category-level transition can be indicative in determining the user's decision process. For example, when a visitor arrives at a new city, he/she is more likely to go to a hotel for a rest than to shop in a mall. Hence a clever system should learn to recommend a hotel list rather than a mall list based on the current context.

Fig. ~\ref{fig:example} shows that a specific user once purchased a $phone\ case$ after he/she bought a $cellphone$ and there is an item-level transition from \emph{cellphone} to \emph{cellphone case}. Such an item dependency  might be not very informative for future prediction when the user buys a \emph{pad}. However, the category level transition from  \emph{electronic} \emph{product} to \emph{electronic} \emph{accessories} will help the system to recommend \emph{pad} \emph{case} to the user.
%where the former falls in the category of case and the latter in electronic product. From the item-level, there is no useful information we can learn about the transaction sequence. But the $electronic\ product\rightarrow case$ sequential rule does exactly exists from the category perspective. Thus a recommender who explores the category context could be more inclined to provide the right answer an $ipad\ case$ noticing the user has bought an $ipad$.
While being aware of the importance of category-level transition, the main challenge is how to capture the  transition patterns in item sequence and those in category sequence independently, and to maintain the relationship between two sequences at the same time.

To tackle these problems, we propose a novel seq2seq translation modeling for context-aware sequential recommendation in this work.
Our idea is inspired by the recent advancement in NMT (neural machine translation). The target of NMT is to translate a source sentence from one language to the target one in another language. NMT not only models the sequential patterns but also captures the semantic relations between two corpora.

Intuitively, we can \emph{treat the item and category sequences as two sentences in different languages} and \emph{model their relations in a translation way}. However, there are still two big gaps between the scenarios in language and recommendation translation. On one hand, a NMT model usually reads the source sentence as a whole then outputs the target. In recommendation, we cannot foresee the future information which means an information leakage and may lead to overfitting. On the other hand, each word from the source corpus is in parallel with one corresponding word in the target corpus. While in our case, a category may contain a number of items which indicates a subsidiary relationship. Hence we further present a \emph{one-by-one strategy} to match  the source and target sequences one by one and design \emph{a variational auto-encoder  as an information filter} to model the subsidiary relationship between the item and category.

To summarize, the major contributions of this work are as follows.
\begin{itemize}
	\item We highlight the importance of exploring the category-level transition patterns  in sequential recommendation. To the best of our knowledge, this has been not exploited in the previous literature.
	\item We develop a basic coupled and an extended tripled seq2seq translation models  to capture item-level and category-level transition patterns independently and maintain the relations between item and category sequences in a translation way.
     \item Extensive experiments over two different public datasets show that the proposed  model significantly outperforms the state-of-the-art methods for the sequential recommendation task.
	
\end{itemize}

\section{Related Work}
This section reviews the literature in sequential recommendation and the related neural translation models.

\subsection{Deep Learning based Sequential Recommendation}
The powerful modeling capacity of deep learning techniques has opened up new opportunities for sequential recommendation whose core task is to predict users' future actions based on previous interactions.

There are a number of pioneer significant works using various deep neural networks and they have shown the improvements over the traditional methods. Recurrent neural network which is originally designed for sequential data, has been widely applied in the sequential recommendation problem. For instance, RNN is employed in DREAM~\cite{yu2016dynamic} to capture the global transition features from users' transaction baskets. Hidasi et al.~\cite{hidasi2015session} propose to apply Gated Recurrent Unit, a variety of RNN, to the session-based recommendation. Despite RNN, convolutional neural network is also adopted in CASER~\cite{tang2018personalized} to deal with union-level skip patterns. Recently, self-attention technique~\cite{vaswani2017attention} is shown to exhibit promising performance in many fields such as computer vision and natural language processing. As a result, several sequential recommendation approaches  such as SASRec~\cite{kang2018self} and ATTRec~\cite{zhang2019next} leverage self-attention mechanism to identify relevant items from history. Other kinds of DNNs like memory network~\cite{chen2018sequential} and gating mechanism~\cite{ma2019hierarchical} are also widely employed in the literature. Generally, these deep learning based methods mainly concentrate on item sequence modeling without considering any context information.

\subsection{Context-aware Sequential Recommendation}
In addition to mining the interaction sequences, researchers are  paying more attentions to additional context information to improve the performance.
Specifically, CA-RNN~\cite{liu2016context} builds two adaptive context matrices to capture the input (weather, category) and transition (distance, time intervals) contexts respectively. Bai et al.~\cite{bai2018attribute} employ an attention mechanism to exploit users' evolving appetite for items' attributes. Zhou et al.~\cite{zhou2018atrank} propose an attention-based framework ATRank which models the users' heterogeneous behaviors.  CSAN \cite{huang2018csan} is proposed based on ATRank to discriminate the significance of individual user behaviors. AIR ~\cite{chen2019air} collectively exploits the rich heterogeneous user interaction actions through the category information. The above methods are all in the E-commerce scenario. In POI recommendation, LBPR \cite{he2017category}, SREM \cite{yao2017serm}, and CAPE \cite{chang2018content} take either the category or textual information into consideration. %There is no denying that great efforts have been paid to incorporate context information for better performance in sequential recommendation.

The aforementioned approaches are all based on the single interaction sequence. Recently, a few multi-sequence based context-aware methods have been proposed. Le et al.~\cite{le2018modeling} develop three twin network structures  to capture the synergies between support (e.g., clicks) and target (e.g., purchases) sequences through fully or partial sharing parameters. STAR~\cite{rakkappan2019context} makes use of stacked RNNs to model item and context sequences, respectively. Overall, capturing  transition patterns in context is attracting more attention. However, none of existing methods models them in an explicit way. Meanwhile, the relation between the interaction sequence and the context sequence has been not exploited in these approaches. In contrast, we propose to model context sequential patterns and maintain relations between category and item at the same time.

\subsection{Neural Machine Translation Methods}
Our method is mainly inspired by the machine translation problem whose target is to translate large amounts of texts from the source languages into ones in the target languages. Benefited from deep neural networks, neural machine translation has achieved great success based on seq2seq architecture ~\cite{sutskever2014sequence} compared with traditional statistical approaches. Self-attention mechanism is also fully utilized to boost the performance~\cite{luong2015effective,bahdanau2014neural}.

There are mainly three components in a NMT model: an encoder for source sentence reading, a decoder for target sentence generating, and a middle-ware for relation modeling. The encoder and decoder are usually modeled with recurrent neural networks and the middle-ware with an attention mechanism. %Besides, replacing RNN with CNN could improve the translation performance according to a relative work~\cite{gehring2016convolutional}.
It is clear that NMT method not only models relations between two corpora but also explores multi-sequential patterns independently which is inherently  identical to  our problem.

\paragraph{Variational Auto-Encoder}
Variational auto-encoder is a widespread generative model proposed by Kingma et al. \shortcite{kingma2013auto}. Many studies have applied VAE to other fields such as recommendation system \cite{liang2018variational,sachdeva2019sequential} and neural machine translation \cite{zhang2016variational}. The powerful generative capacity of VAE enables a method to go beyond the limited modeling ability of linear factor models. Due to the sampling process, incorporating variational auto-encoder can improve the robustness of previous deep neural network based model and boost the performance. More recently, an extended version $\beta$-VAE~\cite{higgins2017beta} is proposed by introducing a hyper-parameter $\beta$  to emphasize the regularization term. It is demonstrated that $\beta$-VAE can learn interpretable factorized latent representations and outperform basic VAE with $\beta\ge 1$. In our work, %we also employ VAE to better capture the relation between the category and the item.
we also take advantage of variational auto-encoder's generative capacity to better capture the relation between the item and category sequences.

\section{Problem Formulation}
In this section, we introduce the problem formulation and then present the preliminary neural machine translation (NMT) Model.

\subsection{Problem Formulation}
Assume that we have a set of users and items denoted by $U=\left\{u_1,u_2,...,u_{|U|}\right\}$ and $I=\left\{i_1,i_2,...,i_{|I|}\right\}$ where $|U|$ and $|I|$ are the numbers of users and items, respectively. For each user $u\in U$, we can obtain a sequence of his/her behaviors sorted by time $s_{ui}=\left\{i^u_1,i^u_2,...,i^u_t\right\}$, where $i^u_t$ denotes the item purchased or rated at time step t by the user $u$.
%Given the sequence of a user $u$'s behaviors $s_u$, we can also find the corresponding sequence of categories denoted by $s_{uc}=\left\{c^u_1,c^u_2,...,c^u_t\right\}$.

In this study, we focus on the problem of context-aware sequential recommendation, where each item $i$ that the user $u$ interacts with at time t may contain various types of additional context information. In our paper, we are mainly interested in the category information for the item and take it as the item's context. Let $C=\left\{c_1,c_2,...,c_{|C|}\right\}$ as the category set, where $|C|$ is the number of categories and each item $i$ always belongs to a specific category $c$. Note that our architecture can be easily extended to model other additional context information.

Given the item sequence $s_{ui}=\left\{i^u_1,i^u_2,...,i^u_t\right\}$ that the user $u$ interacts in the history, and the corresponding category sequence $s_{uc}=\left\{c^u_1,c^u_2,...,c^u_t\right\}$, the goal of context-aware sequential recommendation is to predict the item $i^u_{t+1}$ that the user $u$ is most likely to interact with at the next $t+1$ time step.
For ease of presentation, we list the notations in this paper in Table ~\ref{tab:notations}.
\begin{table}
	\centering
	\small
	\setlength\tabcolsep{3.2pt}
	\caption{List of notations.}
	\vspace{-3mm}
	\label{tab:notations}
	\begin{tabular}{c|p{7.5cm}<{\centering}}
		\hline
		Notation   &Description \\
		\hline
		$U$   &a set of all users \\
		\hline
		$I$   &a set of all items \\
		\hline
		$C$   &a set of all categories \\
		\hline
		$\bm{u}$   &the embedding of a user $u$ \\
		\hline
		$\bm{i}$   &the embedding of an item $i$ \\
		\hline
		$\bm{c}$   &the embedding of a category $c$ \\
		\hline
		$d$   &the dimension of embeddings and the hidden states \\
		\hline
		$\bm{h}$   &the hidden state vector \\
		\hline
		$\bm{W}$   &the weight matrix learned during training \\
		\hline
%\begin{comment}
%		$KL$   &the KL divergence term of the VAE loss function \\
%		\hline
%		$\lambda$   &the hyperparameter for balancing the KL term and prediction term\\
%        \hline
%        $\L$   &the hyperparameter for sequence length\\
%		\hline
%\end{comment}
	\end{tabular}
	\vspace{-3mm}
\end{table}

\subsection{Preliminary: Neural Machine Translation (NMT) Model}
\begin{figure}[htb]
	\centering
	\vspace{-0.2cm}
	\hspace{-0.35cm}\includegraphics[width=0.4\textwidth,height=5.0cm]{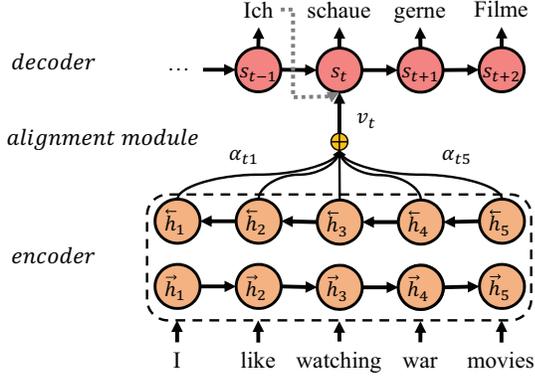}
	\vspace{-0.3cm}
	\caption{Neural Machine Translation (NMT) Model}\label{fig:CNMT}
\end{figure}

The main target of machine translation is to read a sentence like $``I\ like\ watching\ war\ movies"$ and produce $``Ich\ schaue\ gerne\ Filme"$ as the output sentence using the English to Czech task as an example which can also be considered as a seq2seq task. The overview of a classic neural machine translation model is shown in Figure~\ref{fig:CNMT}.

As shown in Fig.~\ref{fig:CNMT}, the NMT framework can be briefly described as an encoder-decoder architecture where the encoder reads the whole source sentence and the decoder translates the outputs of encoder to the target sentence. For encoder, previous works usually employ a bidirectional recurrent neural network that reads the whole source sentence $(\bm{x_1},\bm{x_2},...,\bm{x_{T_x}})$ in a context-aware way, where $\bm{x_n}$ is the embedding of word $x_n$. It will output a sequence of hidden states $(\bm{h_1}, \bm{h_2},...,\bm{h_{T_x}})$ where $\bm{h_t}$ contains context information around word $x_t$ and can be calculated as:
\begin{equation}
\overrightarrow{\bm{h_t}}=\overrightarrow{{\rm RNN}}(\bm{h_{t-1}},\bm{x_t}),
\end{equation}
\begin{equation}
\overleftarrow{\bm{h_t}}=\overleftarrow{{\rm RNN}}(\bm{h_{t+1}},\bm{x_t}),
\end{equation}
\begin{equation}
\bm{h_t}=\overrightarrow{\bm{h_t}}\oplus \overleftarrow{\bm{h_t}},
\end{equation}
where $\oplus$ denotes vector concatenation.

There is an alignment module or so called attention module deciding which part of the source sentence should be focused on. The alignment module is based on the current decoding stage and will derive  a context vector $\bm{v}$:
%Before decoding, the model is expected to decide which part of the source sentence attention should be focused on according to the current decoding stage. Hence, there is always an alignment model or so called attention model deriving a context vector $\bm{c}$ considering all hidden states of the source sentence:
\begin{equation}
\bm{v_{(t-1)j}}=\sum_{j=1}^{T_x}\alpha_{(t-1)j} \bm{h_j},
\end{equation}
\begin{equation}\label{eq:alignment}
\alpha_{(t-1)j}=\frac{{\rm exp}(e_{(t-1)j})}{\sum_{k=1}^{T_x}{\rm exp}(e_{(t-1)k})},
\end{equation}
where $\alpha$ is the weight parameter, and $e_{(t-1)j}$ measures how well the decoding stage at position t-1 and the input hidden state at position j match, and it can be formulated as:
\begin{equation}
e_{(t-1)j}=p(\bm{s_{t-1}},\bm{h_j}),
\end{equation}
where $\bm{s_{t-1}}$ is the hidden state in the decoder which will be defined later, and $p$ is the function that can be defined in many ways such as a multi-layer perceptron.

After the alignment procedure, the decoder starts to generate the target word given the previous predicted words $(y_1,y_2,...,y_{t-1})$ and the current context vector $\bm{v_t}$ commonly through a directional recurrent neural network. It will first calculate the hidden state $\bm{s_{t}}$ of the decoding stage at the time t:
\begin{equation}
\bm{s_t}={\rm RNN}(\bm{s_{t-1}},\bm{v_t},\bm{y_{t-1}}),
\end{equation}
where $\bm{y_{t-1}}$ is the  vector for the (t-1)-th word $y_{t-1}$.

Then the probability of choosing  the next word $y_t$ is generally defined as:
\begin{equation}
\small
p(y_t|y_{<t};(x_1,x_2,...,x_{T_x}))=f(\bm{s_t},\bm{v_t},\bm{y_{t-1}}),
\end{equation}
where the function $f$ aims to fuse information from three sources including the current  hidden state $\bm{s_t}$, the current context vector $\bm{v_t}$, and the previous word vector $\bm{y_{t-1}}$, and gives the final probability score.

Overall, the basic NMT method can model two sequences independently and builds a bridge between them, showing that it can capture the sequential patterns in the source and target sentences as well as  reflect the relation between two sentences. Based on this observation, we introduce the seq2seq language model to the sequential recommendation field.

\section{The Proposed Model}
In this section, We elaborate the proposed architecture in detail. We first present a basic coupled seq2seq translation model (CSTM), and then extend it to a tripled seq2seq translation model (TSTM).

\subsection{Coupled Seq2seq Translation Model}
The key idea of our modeling is to treat the category and item sequences as two sentences to be translated. Different from previous approaches, we aim to extract two types of transition patterns from item and category sequences independently while keeping relations between these two sequences. Taking the example in Fig.~\ref{fig:example} for illustration, the item-level transition $cellphone$ $\rightarrow$ $cellphone case$ and the category-level transition $electronic$ $product$ $\rightarrow$ $electronic$ $accessories$ are two different but inherently associated patterns. When the user later buys a pad in $electronic$ $product$, it would be indicative to first choose the correct category $electronic$ $accessories$ and then recommend the right item $pad$ $case$.

%Thus the relation between above two level patterns should also be maintained due to a transfer action during the user's decision process. Thank to the neural machine translation method which totally meets our need.
Inspired by the NMT mechanism, we propose to model the items and categories as two sentences in two corpus, where the item-level and category-level transition  patterns are encoded in two word sequences, and the relation between item and category sequence is captured by the translation process.

\begin{figure}[htb]
	\centering
	\vspace{-0.2cm}
	\hspace{-0.35cm}\includegraphics[width=0.48\textwidth,height=3.6cm]{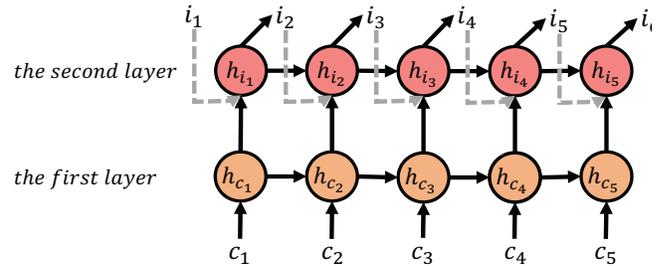}
	\vspace{-0.2cm}
	\caption{Coupled Seq2seq Translation Model (CSTM)}\label{fig:CT}
\end{figure}

In this section, we first present a basic coupled seq2seq translation model (CSTM) which directly translates category sequence into item sequence.
CSTM aims to reflect the scenario where the users first decide the item category and then choose the item. For example, when a user wants to buy some fruits, he/she will browse the webpages containing various fruits like oranges and apples and then make the selection. The architecture of CSTM is shown in Figure~\ref{fig:CT}.

Similar to the NMT method, there are two main components in CSTM: the first layer and the second layer. Please note that we use the $first$, $second$, and $third$ layer instead of encoder and decoder to refer to different parts in our model because we will introduce multi-layers in the following section.

Different from the bidirectional RNN in NMT, we propose a \emph{one-by-one strategy} to translate each category into an item instead of operating on whole sentence. The details are as follows.
In the first layer (encoder) in CSTM in Fig. ~\ref{fig:CT}, we apply a single forward RNN, rather than the bi-directional RNN in NMT, to the input category sequence. In the language translation scenario, the bi-directional RNN are beneficial to the target sentence generation. By reading the whole source sentence in a back and forward way, it increases the awareness of the main intention.
%Given a word from the source sentence, the language translation task tries to find a word in the target corpus with the same meaning such as $movie$ to $Filme$ and $I$ to $Ich$ where the source word is already known.
However, the  nature of sequential recommendation task is quite different.  When we are going to predict which item the user will interact at the next step, we cannot be informed of the next category in advance. Hence we should not model the category sequence with bi-directional RNN which encodes the information of the whole source sentence.
%Instead, we should predict the user's future action at both item and category level.

In the second layer (decoder) in CSTM in Fig. ~\ref{fig:CT}, given the fact that we know the first category $c_1$ and its corresponding item $i_1$, our goal is to predict the next item $i_2$. This is also different from that in NMT which tries to find a word in the target corpus with the same meaning, e.g., $I$ to $Ich$, and $movie$ to $Filme$, showing a mapping from $(s_1,s_2,s_3,s_4,s_5)$ to $(t_1,t_2,t_3,t_4,t_5)$.
In contrast, in our case, we are going to translate the sequence $(c_1,c_2,c_3,c_4,c_5)$ into $(i_2,i_3,i_4,i_5,i_6)$.
%We call this as the $dislocation$ issue.

%We now propose to adapt NMT to address the $dislocation$ issue.
We now present the detail in adapting NMT to our next item prediction problem.
Specifically, in Fig. ~\ref{fig:CT}, the first layer takes a sequence of category $(\bm{c_1},\bm{c_2}$,...,$\bm{c_t})$ as the input, where $\bm{c_i}\in \mathbb{R}^{d\times 1}$ is the embedding of the i-th category $c_i$ and $d$ is the dimension size of embedding. The output is a hidden category state sequence $(\bm{h_{c1}},\bm{h_{c2}}$,...,$\bm{h_{ct}})$ where each $\bm{h_{c_t}}\in \mathbb{R}^{d\times1}$ is calculated as:
\begin{equation}
\bm{h_{c_t}}={\rm RNN}(\bm{h_{c_{t-1}}},\bm{c_t})
\end{equation}

%For instance, when predicting whether a user $u$ will buy a keyboard or not, we should first foresee the correct category.

We then input the hidden category states into the second layer directly, without going through the commonly used alignment module in NMT. As we have analyzed before, we should not get any explicit clues from future, whereas the output of the alignment module will contain information about the whole input sentence, and this may lead to information leakage and overfitting.

%Remember that in Eq. ~\ref\{eq:alignment} of the NMT method, the alignment module will output a context vector $\bm{v_t}$.

Another issue in the second layer is that we would like to take the category information into account due to the correlation between the category and the item. From the category perspective, $\bm{h_{c_t}}$ is the category state mostly relevant to the next item $i_{t+1}$. Thus we define the hidden item state $\bm{h_{i_t}}\in \mathbb{R}^{d\times1}$  in the second layer in CSTM as follows:
\begin{equation}\label{eq:hit}
\bm{h_{i_t}}={\rm RNN}(\bm{h_{i_{t-1}}},\bm{i_t},\bm{h_{c_{t}}}),
\end{equation}
where  $\bm{h_{i_{t-1}}} \in \mathbb{R}^{d\times1}$ is the (t-1)-th hidden item state, $\bm{i_t}\in \mathbb{R}^{d\times1}$ is the embedding for the t-th item, $\bm{h_{c_{t}}} \in \mathbb{R}^{d\times1}$ is the t-th hidden category state.
%$\bm{h_{i_t}}\in \mathbb{R}^{d\times1}$ is utilized for next item $i_{t+1}$ prediction.

To train the model, we maximize the point-wise ranking loss function at the step t which can be formulated as the log likelihood:
\begin{equation}
\mathcal L=\mathcal L_1 + \mathcal L_2
\end{equation}
%\begin{equation}\label{eq:L1}
%\mathcal L_1={\rm log}(p(i_{t+1}|i_{\leq {t}},c_{\leq {t}}))
%\end{equation}
%\begin{equation}\label{eq:L2}
%\mathcal L_2={\rm log}(p(c_{t+1}|c_{\leq {t}}))
%\end{equation}
\begin{equation}\label{eq:L1}
\mathcal L_1={\rm log}(p(\bm{i_{t+1}}|\bm{h_{i_t}})),
\end{equation}
\begin{equation}\label{eq:L2}
\mathcal L_2={\rm log}(p(\bm{c_{t+1}}|\bm{h_{c_t}})),
\end{equation}
The loss $\mathcal L$ at step t consists of two terms: $\mathcal L_1$ for the next item prediction and  $\mathcal L_2$ for the next category prediction. The probability $p(\bm{i_{t+1}}|\bm{h_{i_t}})$ and $p(\bm{c_{t+1}}|\bm{h_{c_t}})$ in Eq. ~\ref{eq:L1} and Eq. ~\ref{eq:L2} can be calculated by adding softmax layers over the hidden category and item state $\bm{h_{i_t}}$ and $\bm{h_{c_t}}$, respectively:
\begin{equation}
p(\bm{i_{t+1}}|\bm{h_{i_t}})={\rm softmax}(\bm{W_1}\bm{h_{i_t}})
\end{equation}
\begin{equation}
p(\bm{c_{t+1}}|\bm{h_{c_t}})={\rm softmax}(\bm{W_2}\bm{h_{c_t}})
\end{equation}
where $\bm{W_1}\in \mathbb{R}^{|I|\times d}$, $\bm{W_2}\in \mathbb{R}^{|C|\times d}$ are trainable parameters. It is worth noting that there is only one loss function for target sentence prediction in original NMT model. Our recommendation problem is much more difficult since the future information is unknown. We add the category loss $\mathcal L_2$ in our case so as to make the proposed method to have the ability to predict the category, which in turn help predict the item.
%It's not contradictory to the considerations mentioned above, because the future information from prediction loss function is very implicit which can further improve the decoder's performance.

\subsection{Tripled Seq2seq Translation Model}
In the previous section, we design a basic CSTM to directly utilize the category-level transition pattern for the prediction of the next item. However, the relation between the item-level and category-level transition patterns is not fully explored due to the one-way information passing from category to item. While treating category as auxiliary information helps item prediction, the item can also assist the category prediction in a reverse way. Inspired by the concept of back-translation \cite{sennrich2015improving}, we further propose an extended tripled seq2seq translation model (TSTM).

The key idea of TSTM is to translate item sequence into category sequence at the first stage and then translate it back to item sequence at the second stage. We believe that item and category sequences can unite tightly and benefit from each other during the generating process. In other words, item sequence can help predict category more precisely and the generated  high-quality category sequence can improve the item prediction.

Intuitively, it would be easy to translate the item sequence into the category sequence. However, due to the subsidiary relationship between category and item, i.e., category is an abstract and high-level description about the specific item, it is hard to translate item sequence $(i_1,i_2,i_3,i_4,i_5)$ into category sequence $(c_2,c_3,c_4,c_5,c_6)$ by directly applying the NMT model, where the source and the target word are of the same semantic level. To address the problem, we introduce a generative \emph{variational auto-encoder \cite{kingma2013auto} (VAE) as an information filter} to model such a subsidiary relationship.

VAEs have been widely used in language modeling and recommendation \cite{zhang2016variational,liang2018variational,sachdeva2019sequential} owing to the power of learning a compressed representation $\bm{z}$ of the input picture or input sequence. The conventional loss function of the VAE ~\cite{kingma2013auto}  is formulated as:
\begin{equation}\label{eq:vae}
	L=-D_{KL}(q(\bm{z}|\bm{x})||p(\bm{z}))+\mathbb{E}_{q(\bm{z}|\bm{x})}[{\rm log}(p(\bm{x}|\bm{z}))],
\end{equation}\label{eq:vae}
where $\bm{x}$ is the input observed data, $\bm{z}$ is the latent factor variable, $q(\bm{z}|\bm{x})$ and $p(\bm{x}|\bm{z})$ are inference and generative model, and $D_{KL}(q||p)$ is the Kullback-Leibler divergence between two distributions. In Eq. \ref{eq:vae}, the expectation part  $\mathbb{E}$ can be viewed as reconstruction loss while the KL part as regularization. Furthermore, $\beta$-VAE introduces a hyper-parameter $\beta$ to push the model to learn a disentangled representation of the data:
\begin{equation}\label{eq:betavae}
	L=-D_{KL}(q(\bm{z}|\bm{x})||p(\bm{z}))+\beta\mathbb{E}_{q(\bm{z}|\bm{x})}[{\rm log}(p(\bm{x}|\bm{z}))]
\end{equation}

\begin{figure}[htb]
	\centering
	\vspace{-0.4cm}
	\hspace{-0.35cm}\includegraphics[width=0.38\textwidth,height=7.2cm]{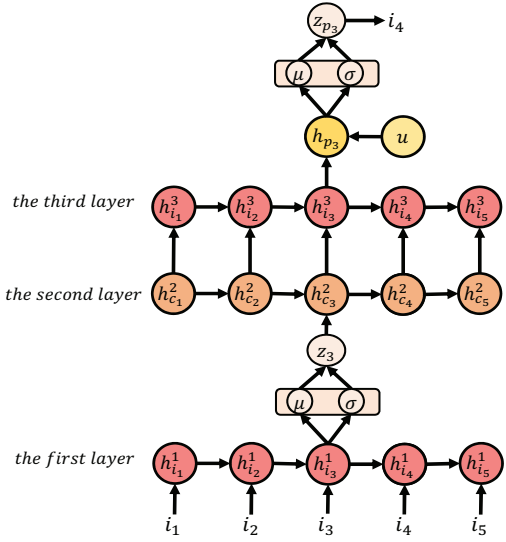}
	\vspace{-0.3cm}
	\caption{Tripled Seq2seq Translation Model (TSTM)}\label{fig:FT}
	\vspace{-0.3cm}
\end{figure}

In order to model the subsidiary relationship between the category and item, we employ a continuous latent variable $\bm{z}$ at each step in item sequence, which will decide what category the next item should be selected from. By introducing the variable $\bm{z}$, we can sample an item which belongs to the same category of the previous item at the next time step. This will help the category prediction more robust. Based on the above assumption, we design our VAE incorporated tripled seq2seq translation model. The architecture of TSTM is shown in Figure~\ref{fig:FT}.

TSTM mainly consists of three layers in Figure~\ref{fig:CT}. The first layer reads the item sequence, then the second layer translates the item into category, and finally the third layer back translates category into item. Before the second layer and the final prediction, there are two VAEs processing the output of the first layer and the third layer, respectively.

Specifically, we feed the item embedding sequence $(\bm{i_1},\bm{i_2},...,\bm{i_t})$ to the first layer and get the output of hidden item state $\bm{h^1_{i_t}}\in \mathbb{R}^{d\times1}$ at step t:
\begin{equation}
	\bm{h^1_{i_t}}={\rm RNN}(\bm{h^1_{i_{t-1}}},\bm{i_t})
\end{equation}
$\bm{h^1_{i_t}}$ is then used for next item prediction in the first layer:
\begin{equation}\label{eq:L1_t}
	\mathcal L_1={\rm log}(p(i_{i+1}|\bm{h^1_{i_t}})),
\end{equation}
which is similar to the loss function in the coupled seq2seq translation model.
Note that in this subsection we use the superscript to denote the output of each layer.

Next, to model the aforementioned subsidiary relationship, we introduce a VAE part between the first and second layers which is different from the basic CSTM. It works as follows. We infer the latent variable $\bm{z_t}\in \mathbb{R}^{d/2\times1}$ at each time step conditioned on previous actions of the sequence which follows the gaussian distribution:
\begin{equation}
	q(\bm{z_t}|\bm{h^1_{i_t}})=\mathcal N(\mu(\bm{h^1_{i_t}}), {\rm diag}\left\{\sigma^2(\bm{h^1_{i_t}})\right\}),
\end{equation}
where $\mu(\bm{h^1_{i_t}})$ and $\sigma^2(\bm{h^1_{i_t}})$ denote the parameters of a gaussian distribution generated from $\bm{h^1_{i_t}}$, and can be simply defined as:
\begin{equation}
	\mu(\bm{h^1_{i_t}})=\bm{h^1_{i_t}}[d/2:]
\end{equation}
\begin{equation}
	\sigma^2(\bm{h^1_{i_t}})={\rm exp}^{\bm{h^1_{i_t}}[:d/2]}
\end{equation}
Then a latent factor $\bm{z_t}$ is sampled from the above posterior inference distribution using the widely used reparameterization trick to avoid the non differential problem~\cite{kingma2013auto}:
\begin{equation}
	\bm{z_t}=\bm{\mu}+\bm{\sigma} \odot \bm{\epsilon},
\end{equation}
where $\bm{\epsilon}\in \mathbb{R}^{d/2\times1}$ is a vector of standard Gaussian noise variables and $\odot$ denotes an element-wise product operation. In the generating process of the hidden category state $\bm{h^2_{c_t}}\in \mathbb{R}^{d\times1}$, $\bm{z_t}$ is incorporated into the second layer:
\begin{equation}
	\bm{h^2_{c_t}}={\rm RNN}(\bm{h^2_{c_{t-1}}},\bm{c_t},\bm{W}\bm{z_{t}})
\end{equation}
where $\bm{W}\in \mathbb{R}^{d\times d/2}$ is a trainable parameter.
%and $\bm{h^2_{c_t}}\in \mathbb{R}^{d\times1}$ is used for next category prediction
Here we modify the standard VAE loss function into:
\begin{equation}\label{eq:vae_1}
	\mathcal L^1_{vae}=KL_c+\mathcal L_2,
\end{equation}
where $KL_c=-D_{KL}(q(\bm{z_t}|\bm{h^1_{i_t}})||p(\bm{z_t}))$ is the regularization term and $\mathcal{L}_2={\rm log}(p(c_{i+1}|\bm{h^2_{c_t}}))$ is used for next category prediction which is very close to Eq. (\ref{eq:L1_t}). Note that $p(\bm{z_t})$ is the prior over the latent variable $\bm{z}$ and is commonly set to a Gaussian distribution.
%Then we calculate the probability of next category $c_{t+1}$ with the following equation:
%\begin{equation}
%\mathcal L_2=-{\rm log}(p(c_{i+1}|\bm{h^2_{c_t}}))
%\end{equation}
%\begin{equation}
%p(c_{t+1}|\bm{z_t})={\rm sofmax}(\bm{W_c}\bm{h^2_{c_t}})
%\end{equation}
%Now we can modify the conventional VAE loss function into:
%\begin{equation}\label{eq:vae_1}
%\mathcal L=-D_{KL}(q(\bm{z_t}|i_{\leq t})||p(\bm{z_t}))+\mathbb{E}_{q(\bm{z_t}|i_{\leq t})}[{\rm log}(p(c_{t+1}|\bm{z_t}))]
%\end{equation}
%as our second layer's loss function where $p(\bm{z_t})$ is the prior over latent variable $\bm{z}$ which is commonly a Gaussian distribution.
%For simplicity, we rewrite $\mathcal L$ in Eq.~\ref{eq:vae_1} into:
%\begin{equation}
%\mathcal L=KL_c+\mathcal L_2
%\end{equation}
%The second term $\mathcal L_2$ can be viewed as a next category prediction function which is very close to Eq.~\ref{eq:L1_t}.

As we can see, there are mainly two differences between our version and the standard VAE in terms of the sequence recommendation case. (1) The latent variable $\bm{z_t}$ is inferred at each time step conditioned on previous actions of the sequence. (2) We predict the next category conditioned on the latent variable $\bm{z_t}$ instead of reconstructing the input item.

After obtaining the hidden category state $\bm{h^2_{c_t}}$ in the second layer, we introduce the third layer for the purpose of back-translation operation and feed $\bm{h^2_{c_t}}$ back to the third layer for generating the hidden item state $\bm{h^3_{i_t}}$ in a similar way:
\begin{equation}
	\bm{h^3_{i_t}}={\rm RNN}(\bm{h^3_{i_{t-1}}},\bm{i_t},\bm{h^2_{c_t}})
\end{equation}

In sequential recommendation, users' general long term preference is also important for next item prediction in addition to sequential patterns. For example, people may choose to visit different POIs even though they arrive at the same places due to their personal interests. Therefore, we propose to incorporate users' individual preference upon the third layer.
For each user $u\in U$, we will allocate a corresponding personal vector $\bm{u}\in\mathbb{R}^{d\times1}$ reflecting $u$'s static preference. After getting a user's representation $\bm{u}$ and the hidden item state $\bm{h^3_{i_t}}$ from the third layer, we produce a fusion vector $\bm{h_{p_t}}\in\mathbb{R}^{d\times1} $ which integrates the user's dynamic and static interests:
\begin{equation}
	\bm{h_{p_t}}=\bm{W_f}(\bm{u}\oplus \bm{h^3_{i_t}}),
\end{equation}
where $\bm{W}\in\mathbb{R}^{2d\times d}$ is a trainable fusion matrix. Then, we employ a VAE again to generate the final representation $\bm{z_{p_t}}\in\mathbb{R}^{d/2\times1}$ for the item prediction due to the same reason as before. We also find that VAE can enhance the robustness of our model for the sample process by introducing noises. If we simply make use of $\bm{h^3_{i_t}}$, our model will gradually pay more attention to the personal vector $\bm{u}$ yet ignoring the effect of seq2seq translation part. At last, we similarly modify Eq. ~\ref{eq:vae}  into:
\begin{equation}\label{eq:vae_2}
	%\mathcal L=-D_{KL}(q(\bm{z_{p_t}}|i_{\leq t})||p(\bm{z_{p_t}}))+\mathbb{E}_{q(\bm{z_{p_t}}|i_{\leq t})}[{\rm log}(p(i_{t+1}|\bm{z_{p_t}}))]
	\mathcal L^2_{vae}=KL_i+\mathcal{L}_3,
\end{equation}
where $KL_i=-D_{KL}(q(\bm{z_{p_t}}|\bm{h_{p_t}})||p(\bm{z_{p_t}}))$ is the K-L divergence between two distributions, and $\mathcal{L}_3={\rm log}(p(i_{t+1}|\bm{z_{p_t}}))$ is introduced for item recommendation considering user's static interest. The approximate posterior $q(\bm{z_{p_t}}|\bm{h_{p_t}})$ is formulated as:
\begin{equation}
	q(\bm{z_{p_t}}|\bm{h_{p_t}})=\mathcal N(\mu(\bm{h_{p_t}}), {\rm diag}\left\{\sigma^2(\bm{h_{p_t}})\right\})
\end{equation}
\begin{equation}
	\mu(\bm{h_{p_t}})=\bm{h_{p_t}}[d/2:]
\end{equation}
\begin{equation}
	\sigma^2(\bm{h_{p_t}})={\rm exp}^{\bm{h_{p_t}}[:d/2]}
\end{equation}
%Also the probability of the next item conditioned on $\bm{z_{p_t}}$ is defined as following:
%\begin{equation}
%p(i_{t+1}|\bm{z_{p_t}})={\rm softmax}(\bm{W_l}\bm{z_{p_t}})
%\end{equation}
%$\bm{W_l}\in\mathbb{R}^{|I|\times d/2}$ is a weight matrix learned during training. For the same reason in the second layer, we rewrite $\mathcal L$ in Eq.~\ref{eq:vae_2} into:
%\begin{equation}
%\mathcal{L}=KL_i+\mathcal{L}_3
%\end{equation}
Finally, to train our model, we define the sum loss function of TSTM considering all factors before as:
\begin{equation}
	\mathcal L_{sum}=\mathcal L_1 + \mathcal L^1_{vae} +\mathcal L^2_{vae}\mathcal = \mathcal L_1 + \mathcal L_2 +\mathcal L_3 +KL_c+KL_i
\end{equation}
%which can be optimized by maximizing $\mathcal L_{sum}$.
Furthermore, we import a hyperparameter $\lambda$ to balance the $KL$ term and the prediction term. Based on the observation of $\beta$-VAE ~\cite{higgins2017beta}, a higher weight on the $KL$ term helps the model to learn disentangled representations of independent data factors and can  improve performance. Thus, we define the final loss objective as:
\begin{equation}
	\mathcal L_{sum}=\mathcal L_1 + \mathcal L_2 +\mathcal L_3 +\lambda (KL_c+KL_i),
\end{equation}
where $\lambda$ is a hyper-parameter and we will examine its effects in the experiment part.

Our TSTM can also be extended to a stacked version S-TSTM  by translating item into category once more based on the top layer's outputs of a single TSTM  and then back to item before the personalized part. Through this way, the translation method can be further enhanced and the relations between two sequences are captured in a more comprehensive way.

\section{Experiments}
In this section, we first give a detailed description of two public datasets from different sources. We then describe the baselines. Finally, we present and analyze the empirical results.

\subsection{Datasets}
We conduct experiments on two datasets from different sources, including MovieLens, Gowalla.

%\begin{itemize}
	%\item
	\textbf{MovieLens} is the widely used benchmark dataset for evaluating recommender systems. We use the MovieLens-1M version \footnote{https://grouplens.org/datasets/movielens/1m/} which contains 1,000,209 ratings by 6,040 users on 3,900 movies from 18 categories such as Action, Comedy, and Romance.

	%\item
	\textbf{Gowalla}  is collected from a real-world location based social network Gowalla. Each record consists of occurrence time, GPS location, and corresponding user ID. The dataset is supplemented with categories by Yang et al. ~\cite{yang2017bridging}. We use this version in our experiment.

%	%\item
%	\textbf{Tmall} is an e-commerce dataset collected from the largest B2C platform in China which contains millions of user transactions \footnote{https://tianchi.aliyun.com/dataset/dataDetail?dataId=47}. The users' interactions are combined with clicks,  bookmarks, and purchase actions. Due to the large scale, we select the interactions in June for training and evaluation.
%There are four types of user behaviors: check, collect, add-to-cart and purchase. Additional information such as category and brand is also released. Due to the large scale, we only select the interactions in June for training and evaluation.
%\end{itemize}

Following the settings in previous studies ~\cite{tang2018personalized,ma2019hierarchical,kang2018self}, we preprocess the above datasets by removing cold-start users and items. For MovieLens, we first treat different types of behaviors equally and convert the explicit actions to implicit feedback of 1. We then remove the inactive users who visit less than $n$ items and unpopular items checked by less than $n$ users, where $n$ is 5, 10 for MovieLens and Gowallal, respectively. Furthermore, we retain users who have more than 20 records on Gowalla to ensure the sequence length ~\cite{li2018next}. For each user, the most recently visited item is considered as the test item while the second one is for validation, and all other items are for training ~\cite{kang2018self}. The statistics of two preprocessed datasets are listed in Table~\ref{tab:Statistics_datasets}.

\begin{table}
\vspace{-2mm}
	\centering
	\small
	\setlength\tabcolsep{3.2pt}
	\caption{Statistics of the evaluation datasets.}
	\vspace{-3mm}
	\label{tab:Statistics_datasets}
	\begin{tabular}{llllll}
		\hline
		Datasets   &\#user &\#item &\#interaction&\#category  &sparsity \\
		\hline
		MovieLens &6,040  & 3,416   & 999,611 &18  &95.16\%\\
		Gowalla &13,989  &22,239  & 896,506 &354  &99.71\%\\	
%		Tmall  &8,597	&9,907	&833,624 &213	&99.02\%\\
		\hline
	\end{tabular}
	\vspace{-3mm}
\end{table}

\subsection{Evaluation Metrics}
We adopt two widely used metrics $Hit@n$ and $NDCG@n$ to evaluate the performance of recommendation models~\cite{kang2018self,he2017neural}. Given a list of top N predicted items for the specific user, if the ground truth item $i$ is in the list then we have $Hit_i@n=1$, otherwise $Hit_i@n=0$. We can compute the $Hit@n$ by:
\begin{equation}
	Hit@n=\frac{1}{|D_{test}|}\sum_{i=1}^{|D_{test}|}Hit_i@n,
\end{equation}
where $|D_{test}|$ is the number of totally test examples.

While $Hit@n$ mainly cares about whether $i$ is among the list, $NDCG@n$ focuses more on the ground truth items' explicit ranking. If an item $i$ is ranked at the $i$-th position among the predicted list, we can calculate $NDCG@n$ by:
\begin{equation}
	NDCG@n=\frac{1}{|D_{test}|}\sum_{i=1}^{|D_{test}|}NDCG_i@n,
\end{equation}
where $NDCG_i@n=log_2{(rank_i+2)}$.

When generating the predicted list, we follow the strategy employed in SASRec and NCF~\cite{kang2018self,he2017neural}, i.e., ranking the test item and other $N$ items randomly sampled from unvisited items by the specific user. In our paper, we set $N=500$ for all recommendation models.

\begin{table*}[ht]
	\vspace{-0mm}
	\centering
	\small
	\caption{Performance comparison on two datasets w.r.t. $Hit@K$ and $NDCG@K$.}
	\vspace{-3mm}
	\label{tab:sum_results}
	\renewcommand{\arraystretch}{0.95}
	\setlength\tabcolsep{2pt}
	\begin{tabular}{c|c|cccccccccc}
		\hline
		Dataset&Method&Hit@1&Hit@5&Hit@10&Hit@15&Hit@20&NDCG@1&NDCG@5&NDCG@10&NDCG@15&NDCG@20 \\
		\hline
		\multirow{11}{*}{MovieLens}&Caser&0.1614&0.4205&0.5575&0.6293&0.6849&0.1614&0.2937&0.3382&0.3572&0.3703 \\
		&SASRec&0.1599&0.4308&\underline{0.5748}&\underline{0.6500}&\underline{0.6980}&0.1599&0.2992&0.3460&0.3660&0.3773 \\
		&HGN&0.1535&0.3762&0.5055&0.5821&0.6321&0.1535&0.2675&0.3095&0.3298&0.3416\\
%\cline{2-12}
		&STAR&0.1483&0.3399&0.4502&0.5185&0.5667&0.1483&0.2471&0.2828&0.3009&0.3123\\
		&STAR-C&0.1611&0.3846&0.4950&0.5626&0.6066&0.1611&0.2760&0.3118&0.3296&0.3400\\
		&ANAM&0.0593&0.2018&0.3118&0.3949&0.4599&0.0593&0.1304&0.1658&0.1878&0.2032\\
		&CBS-SN&0.1960&\underline{0.4449}&0.5644&0.6263&0.6674&0.1960&0.3253&0.3640&0.3803&0.3900\\
		&CBS-CFN&\underline{0.2012}&\underline{0.4449}&0.5674&0.6320&0.6714&\underline{0.2012}&\underline{0.3272}&\underline{0.3669}&\underline{0.3840}&\underline{0.3933}\\
		&CBS-DFN&0.1594&0.3623&0.4742&0.5445&0.5954&0.1594&0.2630&0.2993&0.3180&0.3300\\
		\cline{2-12}
		&TSTM&0.2116&\textbf{0.4672}&0.5861&\textbf{0.6550}&\textbf{0.7013}&0.2116&0.3451&0.3836&\textbf{0.4019}&\textbf{0.4129}\\
		&S-TSTM&\textbf{0.2147}&0.4669&\textbf{0.5879}&0.6508&0.6955&\textbf{0.2147}&\textbf{0.3455}&\textbf{0.3848}&0.4015&0.4120\\
		
		\hline
		\multirow{11}{*}{Gowalla}&Caser&0.3081&0.6136&0.7505&0.8171&0.8530&0.3081&0.4680&0.5126&0.5302&0.5387 \\
		&SASRec&0.4140&\underline{0.7002}&\underline{0.8023}&\underline{0.8510}&\underline{0.8818}&0.4140&\underline{0.5645}&\underline{0.5978}&\underline{0.6107}&\underline{0.6180} \\
		&HGN&0.3259&0.6069&0.7276&0.7924&0.8303&0.3257&0.4732&0.5123&0.5295&0.5385\\
%\cline{2-12}
		&STAR&0.2634&0.4867&0.5915&0.6525&0.6930&0.2634&0.3805&0.4144&0.4305&0.4401\\
		&STAR-C&0.2113&0.4249&0.5346&0.6045&0.6524&0.2113&0.3220&0.3575&0.3760&0.3874\\
		&ANAM&\underline{0.4588}&0.6307&0.7165&0.7646&0.8006&\underline{0.4588}&0.5482&0.5760&0.5887&0.5972\\
		&CBS-SN&0.3973&0.6334&0.7353&0.7898&0.8294&0.3973&0.5223&0.5553&0.5698&0.5791\\
		&CBS-CFN&0.4018&0.6401&0.7425&0.7950&0.8301&0.4018&0.5280&0.5614&0.5753&0.5836\\
		&CBS-DFN&0.3748&0.6080&0.7096&0.7660&0.8026&0.3748&0.4969&0.5298&0.5447&0.5534\\
		\cline{2-12}
		&TSTM&0.4747&0.7124&0.8086&0.8526&0.8820&0.4747&0.6003&0.6314&0.6431&0.6500\\
		&S-TSTM&\textbf{0.4762}&\textbf{0.7160}&\textbf{0.8092}&\textbf{0.8540}&\textbf{0.8822}&\textbf{0.4762}&\textbf{0.6036}&\textbf{0.6339}&\textbf{0.6457}&\textbf{0.6524}\\
		
%		\hline
%		\multirow{11}{*}{Tmall}&Caser&0.3159&\textbf{0.5705}&\textbf{0.6596}&\underline{0.7012}&\underline{0.7303}&0.3159&0.4518&0.4806&0.4916&0.4985 \\
%		&SASRec&0.3308&0.5553&0.6511&\textbf{0.7028}&\textbf{0.7337}&0.3308&0.4490&0.4800&0.4937&0.5011 \\
%		&HGN&0.3329&0.5372&0.6190&0.6612&0.6913&0.3329&0.4407&0.4672&0.4784&0.4855\\
%%\cline{2-12}
%		&STAR&0.2402&0.3971&0.4676&0.5125&0.5457&0.2402&0.3218&0.3446&0.3564&0.3643\\
%		&STAR-C&0.2636&0.4399&0.5206&0.5695&0.6031&0.2636&0.3564&0.3822&0.3952&0.4031\\
%		&ANAM&0.3623&0.4898&0.5534&0.5975&0.6274&0.3623&0.4283&0.4489&0.4605&0.4676\\
%		&CBS-SN&\underline{0.3777}&0.5446&0.6167&0.6588&0.6908&\underline{0.3777}&\underline{0.4664}&\underline{0.4897}&\underline{0.5008}&\underline{0.5084}\\
%		&CBS-CFN&0.3739&0.5462&0.6207&0.6616&0.6923&0.3739&0.4649&0.4890&0.4998&0.5070\\
%		&CBS-DFN&0.2872&0.4602&0.5346&0.5793&0.6160&0.2872&0.3777&0.4018&0.4136&0.4223\\
%		\cline{2-12}
%		&TSTM&0.3734&0.5583&0.6373&0.6827&0.7161&0.3734&0.4711&0.4967&0.5086&0.5165\\
%		&S-TSTM&\textbf{0.3779}&\underline{0.5626}&\underline{0.6408}&0.6908&0.7182&\textbf{0.3779}&\textbf{0.4746}&\textbf{0.4998}&\textbf{0.5130}&\textbf{0.5195}\\
		\hline
	\end{tabular}
	%		\vspace{-3mm}
\end{table*}

\subsection{Baseline Methods}
We compare our model with the following state-of-art sequential recommendation methods.

\emph{the state-of-art baselines without considering context information}:
\begin{itemize}
	\item
	\textbf{Caser} is a convolutional sequence embedding recommendation model~\cite{tang2018personalized} which utilizes CNN to capture point- and union-level personalized transition patterns for the Top-k sequential recommendation.
	\item
	\textbf{SASRec} ~\cite{kang2018self} is an application of self-attention mechanism for sequential recommendation problem in order to identify the relevant items from historical records and use them for prediction.
	\item
	\textbf{HGN} (hierarchical gating network) ~\cite{ma2019hierarchical} is a recent proposed architecture for next item recommendation which consists of three modules: feature gating, instant gating, and item-item product. It is designed to capture users' long and short-term preferences.
\end{itemize}

\emph{the state-of-art context-aware baselines}:
\begin{itemize}
	\item
	\textbf{STAR} ~\cite{rakkappan2019context} evolved from CA-RNN~\cite{liu2016context} based on stacked recurrent neural network. Note that this method considers context information as a independent sequence.
	\item
	\textbf{STAR-C} The original STAR only utilizes the temporal context for sequence modeling. We implement a category version of STAR which exploits category context in the same way for a fair comparison.
	\item
	\textbf{ANAM} employs a hierarchical attentive RNN to track the users' evolving appetite for items dynamically~\cite{bai2018attribute}.
	\item
	\textbf{CBS} models a pair of contemporaneous sequences with a twin network ~\cite{le2018modeling}. It predicts the next item in the target sequence (e.g., purchases) with the assistance of a support sequence (e.g., clicks, bookmarks) by fully or partially sharing parameters of two sequence networks. For a fair comparison, We build a support sequence with category and conduct experiments on three variant models including \textbf{CBS-SN} (fully sharing),  \textbf{CBS-CFN} (no sharing),  and \textbf{CBS-DFN} (partially sharing).
\end{itemize}

\emph{our proposed methods}:
\begin{itemize}
	\item
	\textbf{TSTM}  is our proposed tripled seq2seq translation method where  item and category sequences are treated independently and the relations are modeled by a translation way.
	\item
	\textbf{S-TSTM}  is a  stacked version of the tripled seq2seq translation model.
\end{itemize}

Among the baselines, Caser, SASRec, HGN concentrate on the item transaction patterns only. STAR makes use of temporal information as context, and CBS is designed for dealing with the user's additional information.  We adapt both STAR and CBS to the category as the item context. All other methods take the category information into consideration.

\subsection{Experimental Settings}
For a fair comparison, we set all methods' hidden latent state and embedding dimensions $d$ to 50. We set other parameters and training settings in the baseline methods to be consistent with those reported in the original papers. Note that the performance of sequential recommendation methods are highly influenced by the maximum sequence length $n$. To balance the performance and computational complexity, we set $n$ to the length longer than 95\% of the users' historical sequences in the dataset. More specifically, $n$ is 550, 200 for MovieLens, Gowalla respectively.

When training our model, we use a $L$ sliding window over the users' history to generate the training sequences. Such a method is also used in Caser ~\cite{tang2018personalized} and HGN ~\cite{ma2019hierarchical}. We set $L$ to 5 and we investigate the effects of varying $L$. We set the batch size to 128 and the learning rate to 0.001.  To avoid overfitting, we add an extra dropout layer over all embeddings and the dropout rate is set to 0.2. The hyperparameter $\lambda$ is set to 1 and 20 for MovieLens, Gowalla respectively. Note that the optimal setting of $\lambda$ is determined by grid search strategy from $\left\{1, 5, 10, 15, 20\right\}$ on the validation set. %We also choose LSTM for each layer which is very common in neural machine translation methods.

\subsection{Performance Comparison}
Table~\ref{tab:sum_results} shows the overall performance comparison of all methods on two different datasets. We highlight the best results in each column in boldface and underline the second best ones. From Table~\ref{tab:sum_results}, we have the following important observations.

Firstly, our model outperforms all baseline methods in in terms of $NDCG@n$ on two datasets, shown the superior ranking effectiveness. The $Hit@n$ scores of our model are also better than baselines in most of cases. For example, our model achieves an 5.0\%  $Hit@5$ and  5.6\%$NDCG@5$  improvement on MovieLens over the best results from the  baselines. On Gowalla, SASRec and ANAM are the best baselines. However, they are both much worse than our proposed models. 
%On Tmall, the $Hit@n$ scores of our methods  are not prominent but still competitive. The reason might be that the main sequence on Tmall consists of users' clicks and bookmarks besides the purchase actions. The click  and  bookmark behaviors are of low costs, and thus may introduce lots of noises into the data. This makes the category as a sequence become less effective.

Secondly, our stacked version S-TSTM is proved to be more valid on Gowalla dataset compared with the original TSTM. This is consistent with our assumption that the generated sequence of good quality can further help the generating process. Note that there is no distinguished difference between TSTM and S-TSTM on MovieLens due to the much smaller number of categories on this dataset. %We believe that the abundant categories can provide  richer category information and samples.

Among the baseline methods without considering any context information, we find that SASRec is shown to be a very strong benchmark. It is even better than the most recent work HGN. This is owing to the powerful modeling capacity of self-attention mechanism in SASRec which can directly learn from historical bahaviors based on the current state. Also, the setting of input length in our paper is more reasonable than that reported in HGN  where the length is always set to a small value of 50 ~\cite{ma2019hierarchical}.

The baseline methods using category information are generally better than those without category information, indicating that the context information helps improve the performance. However, there are some exceptions. For instance, the performance of context-aware method STAR is inferior since there is no carefully designed sequential pattern capturing module~\cite{rakkappan2019context}. %However, STAR-C can perform better than STAR  except on Gowalla dataset owing to complex category transition patterns in trajectory kind of datasets.

It is also worth mentioning that the CBS framework, which also models the additional context with a separate sequence,  can produce the second best results  in many cases. This strongly demonstrates the effectiveness of capturing the dependency in contexts. Moreover, the superior performance of our model over CBS proves the importance of modeling the relation between the item sequence and the category sequence.

%From a overall perspective, no-sharing method \textbf{CBS-CFN} always performs the best. It reflects the fact that treating the item and category sequences detached exactly improves performance.

\subsection{Ablation Study}
We  conduct a set of ablation experiments to further prove the effectiveness of the proposed translation method.
\subsubsection{Comparison on simple translation methods}
We first introduce three simple but intuitive methods  as follows to see whether the $translation$ idea works or not.
\begin{itemize}
	\item
	\textbf{LSTM:} A simple recurrent neural network for next item prediction. It adopts the LSTM structure to model the item sequence.
	\item
	\textbf{ci Translation (CSTM):} The basic coupled seq2seq translation model aims to translate category into item.
	\item
	\textbf{ici Translation:} A simplified version of our tripled seq2seq translation model (TSTM) without any VAE and personalized functions. This variant is used to fairly compare the two-layer architecture of CSTM and the three-layer architecture of TSTM.
\end{itemize}

To reduce computational complexity, we set the input history sequence length to 200 and 50 for MovieLens and Gowalla datasets. The results are shown in Table~\ref{tab:ablation_1}. It is clear that the LSTM model without any additional information always performs the worst. The ci Translation can outperform LSTM with the assistance of category-level sequential patterns. Moreover, ici Translation achieves the best results in almost all cases. This proves that the effectiveness by highlighting the relations between item and category sequences through a two-way translation.
%
%As for the ci and ici Translation methods, they have the ability to build relationship between category and item through seq2seq way resulting in better performance. We can learn the sequential patterns of items is well exploited with help of category transition patterns, at the same time, the latter can also improve the former by back-translation operation. Thus the translation method is proved to be effective in modeling relationship between item and category sequences.
\begin{table}[ht]
	\vspace{-0mm}
	\centering
	\small
	\caption{The comparison results for simple translation methods.}
	\vspace{-3mm}
	\label{tab:ablation_1}
	\renewcommand{\arraystretch}{0.95}
	\setlength\tabcolsep{2pt}
	\begin{tabular}{|c|c|p{1.5cm}<{\centering}|c|c|}
		\hline
		Dataset&Method&LSTM&ci Translation&ici Translation\\
		\hline
		\multirow{6}{*}{MovieLens}&Hit@5&0.4250&0.4313&\textbf{0.4407} \\
		&Hit@10&0.5374&0.5483&\textbf{0.5522} \\
		&Hit@20&0.6495&0.6596&\textbf{0.6639}\\
		
		&NDCG@5&0.3171&0.3190&\textbf{0.3290}\\
		&NDCG@10&0.3535&0.3570&\textbf{0.3652}\\
		&NDCG@20&0.3819&0.3852&\textbf{0.3935}\\

		\hline
		\multirow{6}{*}{Gowalla}&Hit@5&0.6181&0.6219&\textbf{0.6257} \\
		&Hit@10&0.7284&0.7244&\textbf{0.7292} \\
		&Hit@20&0.8189&0.8164&\textbf{0.8214}\\
		
		&NDCG@5&0.4971&0.5033&\textbf{0.5057}\\
		&NDCG@10&0.5328&0.5366&\textbf{0.5392}\\
		&NDCG@20&0.5557&0.5599&\textbf{0.5626}\\
		
%		\hline
%		\multirow{6}{*}{Tmall}&Hit@5&0.5230&0.5497&\textbf{0.5635} \\
%		&Hit@10&0.6001&\textbf{0.6272}&0.6255 \\
%		&Hit@20&0.6688&\textbf{0.6969}&0.6916\\
%		
%		&NDCG@5&0.4385&0.4629&\textbf{0.5023}\\
%		&NDCG@10&0.4636&0.4880&\textbf{0.5224}\\
%		&NDCG@20&0.4809&0.5057&\textbf{0.5392}\\
		\hline
	\end{tabular}
	%		\vspace{-3mm}
\end{table}

\subsubsection{The effects of variational auto-encoder}
Though the results in Table~\ref{tab:ablation_1} have shown the effectiveness of translating item into category and then back to item. The subsidiary relation  between category and item has not be fully explored without the VAE module. To demonstrate the effectiveness of the VAE structure in our model, we design three $category\ prediction$ variants. The comparison results of these three variants are shown in Table~\ref{tab:ablation_2}.
\begin{itemize}
	\item
	\textbf{LSTM:} A simple recurrent neural network which applies LSTM to the category sequence for next category prediction.
	\item
	\textbf{ic Translation:} A basic coupled seq2seq translation model which translates item into category. It is a inverse version of the coupled seq2seq translation model (CSTM)).
	\item
	\textbf{ivaec Translation:}  An advanced version of ic Translation method. It imports a VAE structure between two layers.
\end{itemize}
\begin{table}[ht]
	\vspace{-0mm}
	\centering
	\small
	\caption{The comparison results for validating the effects of variational auto-encoder.}
	\vspace{-3mm}
	\label{tab:ablation_2}
	\renewcommand{\arraystretch}{0.95}
	\setlength\tabcolsep{2pt}
	\begin{tabular}{|c|c|p{1.5cm}<{\centering}|c|c|}
		\hline
		Dataset&Method&LSTM&ic Translation&ivaec Translation\\
		\hline
		\multirow{6}{*}{MovieLens}&Hit@5&0.8209&\textbf{0.8364}&0.8331 \\
		&Hit@10&0.9430&0.9485&\textbf{0.9488} \\
		&Hit@20&\textbf{1.0000}&\textbf{1.0000}&\textbf{1.0000}\\
		
		&NDCG@5&0.6095&0.6275&\textbf{0.6296}\\
		&NDCG@10&0.6493&0.6644&\textbf{0.6674}\\
		&NDCG@20&0.6641&0.6779&\textbf{0.6808}\\

		\hline
		\multirow{6}{*}{Gowalla}&Hit@5&0.3804&0.3687&\textbf{0.3890} \\
		&Hit@10&0.4878&0.4821&\textbf{0.5055} \\
		&Hit@20&0.6178&0.6218&\textbf{0.6382}\\
		
		&NDCG@5&0.2857&0.2731&\textbf{0.2881}\\
		&NDCG@10&0.3204&0.3097&\textbf{0.3257}\\
		&NDCG@20&0.3531&0.3449&\textbf{0.3592}\\
		
%		\hline
%		\multirow{6}{*}{Tmall}&Hit@5&0.7864&0.7813&\textbf{0.7904} \\
%		&Hit@10&\textbf{0.8771}&0.8719&0.8759 \\
%		&Hit@20&0.9293&0.9289&\textbf{0.9314}\\
%		
%		&NDCG@5&0.6319&0.6296&\textbf{0.6338}\\
%		&NDCG@10&0.6613&0.6592&\textbf{0.6616}\\
%		&NDCG@20&0.6747&0.6738&\textbf{0.6758}\\
		\hline
	\end{tabular}
	%		\vspace{-3mm}
\end{table}

Comparing ic Translation with simple LSTM, we find that item sequence can not always help predict the next category especially on Gowalla dataset. It is most likely due to the subsidiary relationship between category and item. Meanwhile, ivaec Translation version gives the best category prediction results among these methods. We argue that it is difficult for the model to generate more abstract object from the lower-level item representation without any process of distillation. Hence, the introducing of the variational auto-encoder is necessary to compress the information from item level and furthermore assists in the category sequence modeling. The results in Table~\ref{tab:ablation_2} have proved that variational auto-encoder can handle the subsidiary relation  as we have expected.

\subsection{Parameter Study}
\subsubsection{Effects of hyperparameter $\lambda$}
\begin{figure}[htb]
	\centering
	\vspace{-0.2cm}
	\begin{tabular}{cc}
		\hspace{-0.4cm}
		\includegraphics[width=0.15\textwidth,height=4.2cm]{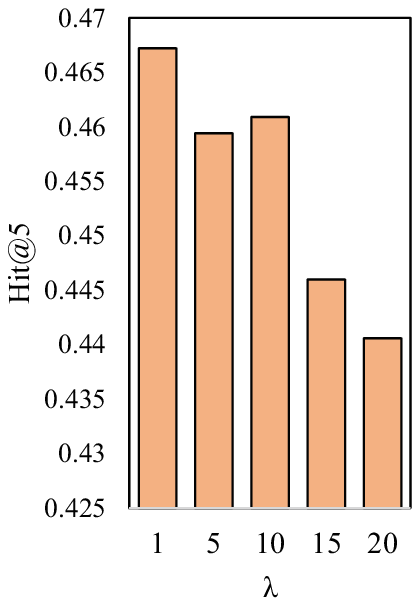}&
		\includegraphics[width=0.15\textwidth,height=4.2cm]{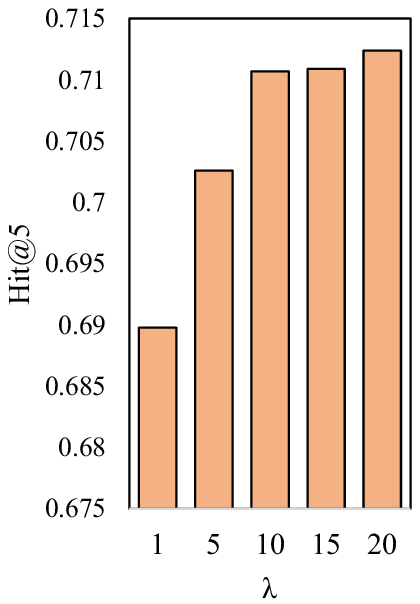}\\
		\quad\quad MovieLens&\quad\quad Gowalla\\
	\end{tabular}
	\vspace{-0.25cm}
	\caption{The $Hit@5$ scores of $\lambda$ varying experiments}\label{fig:lambda_1}
\end{figure}

\begin{figure}[htb]
	\centering
	\vspace{-0.2cm}
	\begin{tabular}{cc}
		\hspace{-0.4cm}\includegraphics[width=0.15\textwidth,height=4.2cm]{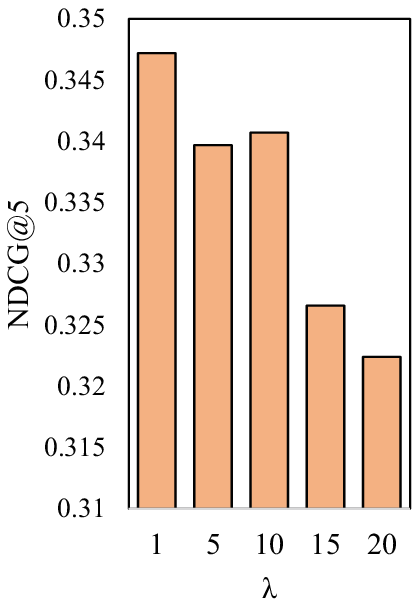} &
		\includegraphics[width=0.15\textwidth,height=4.2cm]{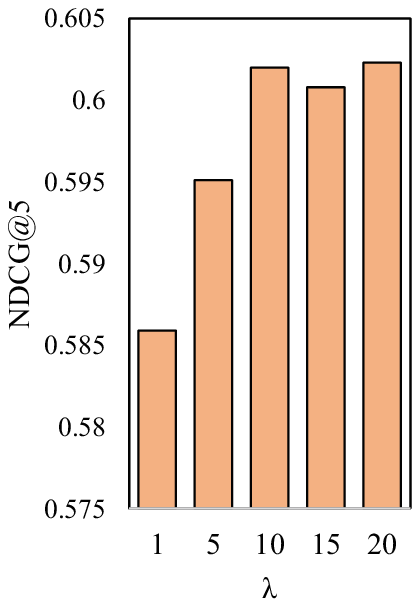}\\
		\quad\quad MovieLens&\quad\quad Gowalla\\
	\end{tabular}
	\vspace{-0.25cm}
	\caption{The $NDCG@5$ scores of $\lambda$ varying experiments}\label{fig:lambda_2}
\end{figure}

We also study the effect of hyperparameter $\lambda$ and give a predetermination principle to approach the best performance of our method. in Figure~\ref{fig:lambda_1} and ~\ref{fig:lambda_2} , The $Hit@5$ and $NDCG@5$ scores of our methods on two datasets are shown by varying $\lambda$ from $\left\{1,5,10,15,20\right\}$. It's obvious that our method achieves the best performance with $\lambda=1$ over MovieLens, while $\lambda=20$ over Gowalla dataset. Note that Gowalla has the largest number of categories while MovieLens the smallest. According to these observations, we infer that a dataset who has a larger number of categories requires a larger $\lambda$. This principle exactly makes sense based on the intuition of $\beta$-VAE~\cite{higgins2017beta}. It has announced that a disentangled representation can be learned with a larger $\lambda$ during training for effective generating. In our method, a larger $\lambda$ means a more strict constrain on the item information before being delivered to the upper category layer. In this way, the most helpful category information could be abstracted from item sequence and hence improves the final category prediction accuracy.

\subsubsection{Effects of hyperparameter $L$}
We study the influence of various slide window length $L$ from $\left\{4, 5, 6, 7, 8\right\}$ and show the results in Table~\ref{tab:ablation_3}. Apparently, setting $L$ too large or too small will both decrease the performance. Different from natural language sentences which always contain a key intention no matter how long the sentence is, we are not sure whether there is one or several intentions within a in sequence.
%Since our translation method performs the best when the sequence expresses one complete intention,
A larger $L$ may introduce noises and give rise to the worse performance. Meanwhile, a smaller $L$ seems to ensure a single one intention but might damage the completeness of sequential patterns. Overall, $L=5$ is likely to be the optimal setting based on the results in Table~\ref{tab:ablation_3}.
\begin{table}[ht]
	\vspace{-0mm}
	\centering
	\small
	\caption{The effect of slide window length $L$.}
	\vspace{-3mm}
	\label{tab:ablation_3}
	\renewcommand{\arraystretch}{0.95}
	\setlength\tabcolsep{2pt}
	\begin{tabular}{|c|p{1.3cm}<{\centering}|p{1.3cm}<{\centering}|c|}
		\hline
		Dataset&$L$&Hit@5&NDCG@5\\
		\hline
		\multirow{6}{*}{MovieLens}&$L$=4&0.4593&0.3414\\
		&$L$=5&0.4672&0.3451 \\
		&$L$=6&\textbf{0.4742}&\textbf{0.3550}\\
		
		&$L$=7&0.4695&0.3496\\
		&$L$=8&0.4675&0.3441\\
		
		\hline
		\multirow{6}{*}{Gowalla}&$L$=4&0.6885&0.5849\\
		&$L$=5&\textbf{0.7124}&\textbf{0.6003}\\
		&$L$=6&0.7117&0.5997\\
		
		&$L$=7&0.6832&0.5771\\
		&$L$=8&0.6787&0.5751\\
		
%		\hline
%		\multirow{6}{*}{Tmall}&$L$=4&0.5544&0.4709 \\
%		&$L$=5&\textbf{0.5583}&\textbf{0.4711} \\
%		&$L$=6&0.5566&0.4703\\
%		
%		&$L$=7&0.5553&0.4673\\
%		&$L$=8&0.5504&0.4677\\
		\hline
	\end{tabular}
	%		\vspace{-3mm}
\end{table}
%\subsubsection{The Study of Dropout Rate}
%We also show that an extra dropout layer can boost the performance.
%
%\begin{figure}[htb]
%	\centering
%	\vspace{-0.2cm}
%	\begin{tabular}{ccc}
%		\hspace{-0.4cm}
%		\includegraphics[width=0.15\textwidth,height=4.2cm]{dropout_ml_Hit.eps}&
%		\includegraphics[width=0.15\textwidth,height=4.2cm]{dropout_gowalla_Hit.eps} & \includegraphics[width=0.15\textwidth,height=4.2cm]{dropout_tmall_Hit.eps}\\
%		\quad\quad MovieLens&\quad\quad Gowalla&\quad\quad Tmall\\
%	\end{tabular}
%	\vspace{-0.25cm}
%	\caption{The $Hit@5$ scores of $\lambda$ varying experiments}\label{fig:dropout_1}
%\end{figure}
%\begin{figure}[htb]
%	\centering
%	\vspace{-0.2cm}
%	\begin{tabular}{ccc}
%		\hspace{-0.4cm}\includegraphics[width=0.15\textwidth,height=4.2cm]{dropout_ml_NDCG.eps} &
%		\includegraphics[width=0.15\textwidth,height=4.2cm]{dropout_gowalla_NDCG.eps} & \includegraphics[width=0.15\textwidth,height=4.2cm]{dropout_tmall_NDCG.eps}\\
%		\quad\quad MovieLens&\quad\quad Gowalla&\quad\quad Tmall\\
%	\end{tabular}
%	\vspace{-0.25cm}
%	\caption{The $NDCG@5$ scores of $\lambda$ varying experiments}\label{fig:dropout_2}
%\end{figure}

\section{Conclusion}
In this paper, we propose a novel translation based architecture for context-aware sequential recommendation which captures the item-level and category-level transition patterns independently while maintaining the relations between these two sequences. To explore the subsidiary relationship between category and item, we further propose to adapt the variational auto-encoder to boosting the performance. Extensive experiments have been conducted over two datasets compared with several state-of-art methods. Results demonstrate the effectiveness of  our proposed method  by the superior performance  over the state-of-the-art baselines.

\section*{Acknowledgments} The work described in this paper has been supported in part by the NSFC projects (61572376, 91646206), and the 111 project(B07037).

%\clearpage

%\bibliographystyle{splncs03}
%\vspace{-1mm}
\bibliographystyle{ACM-Reference-Format}
\bibliography{RecNextReferences}


\begin{thebibliography}{36}

%%% the Web address.

\ifx \showCODEN    \undefined \def \showCODEN     #1{\unskip}     \fi
\ifx \showDOI      \undefined \def \showDOI       #1{#1}\fi
\ifx \showISBNx    \undefined \def \showISBNx     #1{\unskip}     \fi
\ifx \showISBNxiii \undefined \def \showISBNxiii  #1{\unskip}     \fi
\ifx \showISSN     \undefined \def \showISSN      #1{\unskip}     \fi
\ifx \showLCCN     \undefined \def \showLCCN      #1{\unskip}     \fi
\ifx \shownote     \undefined \def \shownote      #1{#1}          \fi
\ifx \showarticletitle \undefined \def \showarticletitle #1{#1}   \fi
\ifx \showURL      \undefined \def \showURL       {\relax}        \fi
% invisible to TeX
\providecommand\bibfield[2]{#2}
\providecommand\bibinfo[2]{#2}
\providecommand\natexlab[1]{#1}
\providecommand\showeprint[2][]{arXiv:#2}

\bibitem[\protect\citeauthoryear{Bahdanau, Cho, and Bengio}{Bahdanau
  et~al\mbox{.}}{2014}]%
        {bahdanau2014neural}
\bibfield{author}{\bibinfo{person}{Dzmitry Bahdanau},
  \bibinfo{person}{Kyunghyun Cho}, {and} \bibinfo{person}{Yoshua Bengio}.}
  \bibinfo{year}{2014}\natexlab{}.
\newblock \showarticletitle{Neural machine translation by jointly learning to
  align and translate}.
\newblock \bibinfo{journal}{\emph{arXiv preprint arXiv:1409.0473}}
  (\bibinfo{year}{2014}).
\newblock


\bibitem[\protect\citeauthoryear{Bai, Nie, Zhao, Zhu, Du, and Wen}{Bai
  et~al\mbox{.}}{2018}]%
        {bai2018attribute}
\bibfield{author}{\bibinfo{person}{Ting Bai}, \bibinfo{person}{Jian-Yun Nie},
  \bibinfo{person}{Wayne~Xin Zhao}, \bibinfo{person}{Yutao Zhu},
  \bibinfo{person}{Pan Du}, {and} \bibinfo{person}{Ji-Rong Wen}.}
  \bibinfo{year}{2018}\natexlab{}.
\newblock \showarticletitle{An attribute-aware neural attentive model for next
  basket recommendation}. In \bibinfo{booktitle}{\emph{The 41st International
  ACM SIGIR Conference on Research \& Development in Information Retrieval}}.
  ACM, \bibinfo{pages}{1201--1204}.
\newblock


\bibitem[\protect\citeauthoryear{Chang, Park, Park, Kim, and Kang}{Chang
  et~al\mbox{.}}{2018}]%
        {chang2018content}
\bibfield{author}{\bibinfo{person}{Buru Chang}, \bibinfo{person}{Yonggyu Park},
  \bibinfo{person}{Donghyeon Park}, \bibinfo{person}{Seongsoon Kim}, {and}
  \bibinfo{person}{Jaewoo Kang}.} \bibinfo{year}{2018}\natexlab{}.
\newblock \showarticletitle{Content-Aware Hierarchical Point-of-Interest
  Embedding Model for Successive POI Recommendation.}. In
  \bibinfo{booktitle}{\emph{IJCAI}}. \bibinfo{pages}{3301--3307}.
\newblock


\bibitem[\protect\citeauthoryear{Chen, Yin, Chen, Yan, Nguyen, and Li}{Chen
  et~al\mbox{.}}{2019}]%
        {chen2019air}
\bibfield{author}{\bibinfo{person}{Tong Chen}, \bibinfo{person}{Hongzhi Yin},
  \bibinfo{person}{Hongxu Chen}, \bibinfo{person}{Rui Yan},
  \bibinfo{person}{Quoc Viet~Hung Nguyen}, {and} \bibinfo{person}{Xue Li}.}
  \bibinfo{year}{2019}\natexlab{}.
\newblock \showarticletitle{AIR: Attentional Intention-Aware Recommender
  Systems}. In \bibinfo{booktitle}{\emph{2019 IEEE 35th International
  Conference on Data Engineering (ICDE)}}. IEEE, \bibinfo{pages}{304--315}.
\newblock


\bibitem[\protect\citeauthoryear{Chen, Xu, Zhang, Tang, Cao, Qin, and Zha}{Chen
  et~al\mbox{.}}{2018}]%
        {chen2018sequential}
\bibfield{author}{\bibinfo{person}{Xu Chen}, \bibinfo{person}{Hongteng Xu},
  \bibinfo{person}{Yongfeng Zhang}, \bibinfo{person}{Jiaxi Tang},
  \bibinfo{person}{Yixin Cao}, \bibinfo{person}{Zheng Qin}, {and}
  \bibinfo{person}{Hongyuan Zha}.} \bibinfo{year}{2018}\natexlab{}.
\newblock \showarticletitle{Sequential recommendation with user memory
  networks}. In \bibinfo{booktitle}{\emph{Proceedings of the eleventh ACM
  international conference on web search and data mining}}. ACM,
  \bibinfo{pages}{108--116}.
\newblock


\bibitem[\protect\citeauthoryear{Chen~Ma and Liu}{Chen~Ma and Liu}{2019}]%
        {ma2019hierarchical}
\bibfield{author}{\bibinfo{person}{Peng~Kang Chen~Ma} {and}
  \bibinfo{person}{Xue Liu}.} \bibinfo{year}{2019}\natexlab{}.
\newblock \showarticletitle{Hierarchical Gating Networks for Sequential
  Recommendation}. In \bibinfo{booktitle}{\emph{Proceedings of the 25th ACM
  SIGKDD International Conference on Knowledge Discovery and Data Mining}}.
  ACM, \bibinfo{pages}{825--833}.
\newblock


\bibitem[\protect\citeauthoryear{He, Li, and Liao}{He et~al\mbox{.}}{2017a}]%
        {he2017category}
\bibfield{author}{\bibinfo{person}{Jing He}, \bibinfo{person}{Xin Li}, {and}
  \bibinfo{person}{Lejian Liao}.} \bibinfo{year}{2017}\natexlab{a}.
\newblock \showarticletitle{Category-aware Next Point-of-Interest
  Recommendation via Listwise Bayesian Personalized Ranking.}. In
  \bibinfo{booktitle}{\emph{IJCAI}}. \bibinfo{pages}{1837--1843}.
\newblock


\bibitem[\protect\citeauthoryear{He, Liao, Zhang, Nie, Hu, and Chua}{He
  et~al\mbox{.}}{2017b}]%
        {he2017neural}
\bibfield{author}{\bibinfo{person}{Xiangnan He}, \bibinfo{person}{Lizi Liao},
  \bibinfo{person}{Hanwang Zhang}, \bibinfo{person}{Liqiang Nie},
  \bibinfo{person}{Xia Hu}, {and} \bibinfo{person}{Tat-Seng Chua}.}
  \bibinfo{year}{2017}\natexlab{b}.
\newblock \showarticletitle{Neural collaborative filtering}. In
  \bibinfo{booktitle}{\emph{Proceedings of the 26th international conference on
  world wide web}}. International World Wide Web Conferences Steering
  Committee, \bibinfo{pages}{173--182}.
\newblock


\bibitem[\protect\citeauthoryear{Hidasi, Karatzoglou, Baltrunas, and
  Tikk}{Hidasi et~al\mbox{.}}{2015}]%
        {hidasi2015session}
\bibfield{author}{\bibinfo{person}{Bal{\'a}zs Hidasi},
  \bibinfo{person}{Alexandros Karatzoglou}, \bibinfo{person}{Linas Baltrunas},
  {and} \bibinfo{person}{Domonkos Tikk}.} \bibinfo{year}{2015}\natexlab{}.
\newblock \showarticletitle{Session-based recommendations with recurrent neural
  networks}.
\newblock \bibinfo{journal}{\emph{arXiv preprint arXiv:1511.06939}}
  (\bibinfo{year}{2015}).
\newblock


\bibitem[\protect\citeauthoryear{Higgins, Matthey, Pal, Burgess, Glorot,
  Botvinick, Mohamed, and Lerchner}{Higgins et~al\mbox{.}}{2017}]%
        {higgins2017beta}
\bibfield{author}{\bibinfo{person}{Irina Higgins}, \bibinfo{person}{Loic
  Matthey}, \bibinfo{person}{Arka Pal}, \bibinfo{person}{Christopher Burgess},
  \bibinfo{person}{Xavier Glorot}, \bibinfo{person}{Matthew Botvinick},
  \bibinfo{person}{Shakir Mohamed}, {and} \bibinfo{person}{Alexander
  Lerchner}.} \bibinfo{year}{2017}\natexlab{}.
\newblock \showarticletitle{beta-VAE: Learning Basic Visual Concepts with a
  Constrained Variational Framework.}
\newblock \bibinfo{journal}{\emph{ICLR}} \bibinfo{volume}{2},
  \bibinfo{number}{5} (\bibinfo{year}{2017}), \bibinfo{pages}{6}.
\newblock


\bibitem[\protect\citeauthoryear{Huang, Zhao, Dou, Wen, and Chang}{Huang
  et~al\mbox{.}}{2018b}]%
        {huang2018improving}
\bibfield{author}{\bibinfo{person}{Jin Huang}, \bibinfo{person}{Wayne~Xin
  Zhao}, \bibinfo{person}{Hongjian Dou}, \bibinfo{person}{Ji-Rong Wen}, {and}
  \bibinfo{person}{Edward~Y Chang}.} \bibinfo{year}{2018}\natexlab{b}.
\newblock \showarticletitle{Improving sequential recommendation with
  knowledge-enhanced memory networks}. In \bibinfo{booktitle}{\emph{The 41st
  International ACM SIGIR Conference on Research \& Development in Information
  Retrieval}}. ACM, \bibinfo{pages}{505--514}.
\newblock


\bibitem[\protect\citeauthoryear{Huang, Qian, Fang, Sang, and Xu}{Huang
  et~al\mbox{.}}{2018a}]%
        {huang2018csan}
\bibfield{author}{\bibinfo{person}{Xiaowen Huang}, \bibinfo{person}{Shengsheng
  Qian}, \bibinfo{person}{Quan Fang}, \bibinfo{person}{Jitao Sang}, {and}
  \bibinfo{person}{Changsheng Xu}.} \bibinfo{year}{2018}\natexlab{a}.
\newblock \showarticletitle{CSAN: Contextual Self-Attention Network for User
  Sequential Recommendation}. In \bibinfo{booktitle}{\emph{2018 ACM Multimedia
  Conference on Multimedia Conference}}. ACM, \bibinfo{pages}{447--455}.
\newblock


\bibitem[\protect\citeauthoryear{Kang and McAuley}{Kang and McAuley}{2018}]%
        {kang2018self}
\bibfield{author}{\bibinfo{person}{Wang-Cheng Kang} {and}
  \bibinfo{person}{Julian McAuley}.} \bibinfo{year}{2018}\natexlab{}.
\newblock \showarticletitle{Self-attentive sequential recommendation}. In
  \bibinfo{booktitle}{\emph{2018 IEEE International Conference on Data Mining
  (ICDM)}}. IEEE, \bibinfo{pages}{197--206}.
\newblock


\bibitem[\protect\citeauthoryear{Kingma and Welling}{Kingma and
  Welling}{2013}]%
        {kingma2013auto}
\bibfield{author}{\bibinfo{person}{Diederik~P Kingma} {and}
  \bibinfo{person}{Max Welling}.} \bibinfo{year}{2013}\natexlab{}.
\newblock \showarticletitle{Auto-encoding variational bayes}.
\newblock \bibinfo{journal}{\emph{arXiv preprint arXiv:1312.6114}}
  (\bibinfo{year}{2013}).
\newblock


\bibitem[\protect\citeauthoryear{Koren, Bell, and Volinsky}{Koren
  et~al\mbox{.}}{2009}]%
        {koren2009matrix}
\bibfield{author}{\bibinfo{person}{Yehuda Koren}, \bibinfo{person}{Robert
  Bell}, {and} \bibinfo{person}{Chris Volinsky}.}
  \bibinfo{year}{2009}\natexlab{}.
\newblock \showarticletitle{Matrix factorization techniques for recommender
  systems}.
\newblock \bibinfo{journal}{\emph{Computer}} \bibinfo{number}{8}
  (\bibinfo{year}{2009}), \bibinfo{pages}{30--37}.
\newblock


\bibitem[\protect\citeauthoryear{LE, LAUW, and Fang}{LE et~al\mbox{.}}{2018}]%
        {le2018modeling}
\bibfield{author}{\bibinfo{person}{Duc~Trong LE}, \bibinfo{person}{Hady~Wirawan
  LAUW}, {and} \bibinfo{person}{Yuan Fang}.} \bibinfo{year}{2018}\natexlab{}.
\newblock \showarticletitle{Modeling contemporaneous basket sequences with twin
  networks for next-item recommendation}. IJCAI.
\newblock


\bibitem[\protect\citeauthoryear{Li, Shen, and Zhu}{Li et~al\mbox{.}}{2018}]%
        {li2018next}
\bibfield{author}{\bibinfo{person}{Ranzhen Li}, \bibinfo{person}{Yanyan Shen},
  {and} \bibinfo{person}{Yanmin Zhu}.} \bibinfo{year}{2018}\natexlab{}.
\newblock \showarticletitle{Next Point-of-Interest Recommendation with Temporal
  and Multi-level Context Attention}. In \bibinfo{booktitle}{\emph{2018 IEEE
  International Conference on Data Mining (ICDM)}}. IEEE,
  \bibinfo{pages}{1110--1115}.
\newblock


\bibitem[\protect\citeauthoryear{Liang, Krishnan, Hoffman, and Jebara}{Liang
  et~al\mbox{.}}{2018}]%
        {liang2018variational}
\bibfield{author}{\bibinfo{person}{Dawen Liang}, \bibinfo{person}{Rahul~G
  Krishnan}, \bibinfo{person}{Matthew~D Hoffman}, {and} \bibinfo{person}{Tony
  Jebara}.} \bibinfo{year}{2018}\natexlab{}.
\newblock \showarticletitle{Variational autoencoders for collaborative
  filtering}. In \bibinfo{booktitle}{\emph{Proceedings of the 2018 World Wide
  Web Conference}}. International World Wide Web Conferences Steering
  Committee, \bibinfo{pages}{689--698}.
\newblock


\bibitem[\protect\citeauthoryear{Liu, Wu, Wang, Li, and Wang}{Liu
  et~al\mbox{.}}{2016b}]%
        {liu2016context}
\bibfield{author}{\bibinfo{person}{Qiang Liu}, \bibinfo{person}{Shu Wu},
  \bibinfo{person}{Diyi Wang}, \bibinfo{person}{Zhaokang Li}, {and}
  \bibinfo{person}{Liang Wang}.} \bibinfo{year}{2016}\natexlab{b}.
\newblock \showarticletitle{Context-aware sequential recommendation}. In
  \bibinfo{booktitle}{\emph{2016 IEEE 16th International Conference on Data
  Mining (ICDM)}}. IEEE, \bibinfo{pages}{1053--1058}.
\newblock


\bibitem[\protect\citeauthoryear{Liu, Wu, Wang, and Tan}{Liu
  et~al\mbox{.}}{2016a}]%
        {liu2016predicting}
\bibfield{author}{\bibinfo{person}{Qiang Liu}, \bibinfo{person}{Shu Wu},
  \bibinfo{person}{Liang Wang}, {and} \bibinfo{person}{Tieniu Tan}.}
  \bibinfo{year}{2016}\natexlab{a}.
\newblock \showarticletitle{Predicting the next location: A recurrent model
  with spatial and temporal contexts}. In \bibinfo{booktitle}{\emph{Thirtieth
  AAAI Conference on Artificial Intelligence}}.
\newblock


\bibitem[\protect\citeauthoryear{Luong, Pham, and Manning}{Luong
  et~al\mbox{.}}{2015}]%
        {luong2015effective}
\bibfield{author}{\bibinfo{person}{Minh-Thang Luong}, \bibinfo{person}{Hieu
  Pham}, {and} \bibinfo{person}{Christopher~D Manning}.}
  \bibinfo{year}{2015}\natexlab{}.
\newblock \showarticletitle{Effective approaches to attention-based neural
  machine translation}.
\newblock \bibinfo{journal}{\emph{arXiv preprint arXiv:1508.04025}}
  (\bibinfo{year}{2015}).
\newblock


\bibitem[\protect\citeauthoryear{Rakkappan and Rajan}{Rakkappan and
  Rajan}{2019}]%
        {rakkappan2019context}
\bibfield{author}{\bibinfo{person}{Lakshmanan Rakkappan} {and}
  \bibinfo{person}{Vaibhav Rajan}.} \bibinfo{year}{2019}\natexlab{}.
\newblock \showarticletitle{Context-Aware Sequential Recommendations
  withStacked Recurrent Neural Networks}. In \bibinfo{booktitle}{\emph{The
  World Wide Web Conference}}. ACM, \bibinfo{pages}{3172--3178}.
\newblock


\bibitem[\protect\citeauthoryear{Rendle, Freudenthaler, Gantner, and
  Schmidt-Thieme}{Rendle et~al\mbox{.}}{2009}]%
        {rendle2009bpr}
\bibfield{author}{\bibinfo{person}{Steffen Rendle}, \bibinfo{person}{Christoph
  Freudenthaler}, \bibinfo{person}{Zeno Gantner}, {and} \bibinfo{person}{Lars
  Schmidt-Thieme}.} \bibinfo{year}{2009}\natexlab{}.
\newblock \showarticletitle{BPR: Bayesian personalized ranking from implicit
  feedback}. In \bibinfo{booktitle}{\emph{Proceedings of the twenty-fifth
  conference on uncertainty in artificial intelligence}}. AUAI Press,
  \bibinfo{pages}{452--461}.
\newblock


\bibitem[\protect\citeauthoryear{Rendle, Freudenthaler, and
  Schmidt-Thieme}{Rendle et~al\mbox{.}}{2010}]%
        {rendle2010factorizing}
\bibfield{author}{\bibinfo{person}{Steffen Rendle}, \bibinfo{person}{Christoph
  Freudenthaler}, {and} \bibinfo{person}{Lars Schmidt-Thieme}.}
  \bibinfo{year}{2010}\natexlab{}.
\newblock \showarticletitle{Factorizing personalized markov chains for
  next-basket recommendation}. In \bibinfo{booktitle}{\emph{Proceedings of the
  19th international conference on World wide web}}. ACM,
  \bibinfo{pages}{811--820}.
\newblock


\bibitem[\protect\citeauthoryear{Sachdeva, Manco, Ritacco, and Pudi}{Sachdeva
  et~al\mbox{.}}{2019}]%
        {sachdeva2019sequential}
\bibfield{author}{\bibinfo{person}{Noveen Sachdeva}, \bibinfo{person}{Giuseppe
  Manco}, \bibinfo{person}{Ettore Ritacco}, {and} \bibinfo{person}{Vikram
  Pudi}.} \bibinfo{year}{2019}\natexlab{}.
\newblock \showarticletitle{Sequential Variational Autoencoders for
  Collaborative Filtering}. In \bibinfo{booktitle}{\emph{Proceedings of the
  Twelfth ACM International Conference on Web Search and Data Mining}}. ACM,
  \bibinfo{pages}{600--608}.
\newblock


\bibitem[\protect\citeauthoryear{Sennrich, Haddow, and Birch}{Sennrich
  et~al\mbox{.}}{2015}]%
        {sennrich2015improving}
\bibfield{author}{\bibinfo{person}{Rico Sennrich}, \bibinfo{person}{Barry
  Haddow}, {and} \bibinfo{person}{Alexandra Birch}.}
  \bibinfo{year}{2015}\natexlab{}.
\newblock \showarticletitle{Improving neural machine translation models with
  monolingual data}.
\newblock \bibinfo{journal}{\emph{arXiv preprint arXiv:1511.06709}}
  (\bibinfo{year}{2015}).
\newblock


\bibitem[\protect\citeauthoryear{Sutskever, Vinyals, and Le}{Sutskever
  et~al\mbox{.}}{2014}]%
        {sutskever2014sequence}
\bibfield{author}{\bibinfo{person}{Ilya Sutskever}, \bibinfo{person}{Oriol
  Vinyals}, {and} \bibinfo{person}{Quoc~V Le}.}
  \bibinfo{year}{2014}\natexlab{}.
\newblock \showarticletitle{Sequence to sequence learning with neural
  networks}. In \bibinfo{booktitle}{\emph{Advances in neural information
  processing systems}}. \bibinfo{pages}{3104--3112}.
\newblock


\bibitem[\protect\citeauthoryear{Tang and Wang}{Tang and Wang}{2018}]%
        {tang2018personalized}
\bibfield{author}{\bibinfo{person}{Jiaxi Tang} {and} \bibinfo{person}{Ke
  Wang}.} \bibinfo{year}{2018}\natexlab{}.
\newblock \showarticletitle{Personalized top-n sequential recommendation via
  convolutional sequence embedding}. In \bibinfo{booktitle}{\emph{Proceedings
  of the Eleventh ACM International Conference on Web Search and Data Mining}}.
  ACM, \bibinfo{pages}{565--573}.
\newblock


\bibitem[\protect\citeauthoryear{Vaswani, Shazeer, Parmar, Uszkoreit, Jones,
  Gomez, Kaiser, and Polosukhin}{Vaswani et~al\mbox{.}}{2017}]%
        {vaswani2017attention}
\bibfield{author}{\bibinfo{person}{Ashish Vaswani}, \bibinfo{person}{Noam
  Shazeer}, \bibinfo{person}{Niki Parmar}, \bibinfo{person}{Jakob Uszkoreit},
  \bibinfo{person}{Llion Jones}, \bibinfo{person}{Aidan~N Gomez},
  \bibinfo{person}{{\L}ukasz Kaiser}, {and} \bibinfo{person}{Illia
  Polosukhin}.} \bibinfo{year}{2017}\natexlab{}.
\newblock \showarticletitle{Attention is all you need}. In
  \bibinfo{booktitle}{\emph{Advances in neural information processing
  systems}}. \bibinfo{pages}{5998--6008}.
\newblock


\bibitem[\protect\citeauthoryear{Wang, Guo, Lan, Xu, Wan, and Cheng}{Wang
  et~al\mbox{.}}{2015}]%
        {wang2015learning}
\bibfield{author}{\bibinfo{person}{Pengfei Wang}, \bibinfo{person}{Jiafeng
  Guo}, \bibinfo{person}{Yanyan Lan}, \bibinfo{person}{Jun Xu},
  \bibinfo{person}{Shengxian Wan}, {and} \bibinfo{person}{Xueqi Cheng}.}
  \bibinfo{year}{2015}\natexlab{}.
\newblock \showarticletitle{Learning hierarchical representation model for
  nextbasket recommendation}. In \bibinfo{booktitle}{\emph{Proceedings of the
  38th International ACM SIGIR conference on Research and Development in
  Information Retrieval}}. ACM, \bibinfo{pages}{403--412}.
\newblock


\bibitem[\protect\citeauthoryear{Yang, Bai, Zhang, Yuan, and Han}{Yang
  et~al\mbox{.}}{2017}]%
        {yang2017bridging}
\bibfield{author}{\bibinfo{person}{Carl Yang}, \bibinfo{person}{Lanxiao Bai},
  \bibinfo{person}{Chao Zhang}, \bibinfo{person}{Quan Yuan}, {and}
  \bibinfo{person}{Jiawei Han}.} \bibinfo{year}{2017}\natexlab{}.
\newblock \showarticletitle{Bridging collaborative filtering and
  semi-supervised learning: a neural approach for poi recommendation}. In
  \bibinfo{booktitle}{\emph{Proceedings of the 23rd ACM SIGKDD International
  Conference on Knowledge Discovery and Data Mining}}. ACM,
  \bibinfo{pages}{1245--1254}.
\newblock


\bibitem[\protect\citeauthoryear{Yao, Zhang, Huang, and Bi}{Yao
  et~al\mbox{.}}{2017}]%
        {yao2017serm}
\bibfield{author}{\bibinfo{person}{Di Yao}, \bibinfo{person}{Chao Zhang},
  \bibinfo{person}{Jianhui Huang}, {and} \bibinfo{person}{Jingping Bi}.}
  \bibinfo{year}{2017}\natexlab{}.
\newblock \showarticletitle{Serm: A recurrent model for next location
  prediction in semantic trajectories}. In
  \bibinfo{booktitle}{\emph{Proceedings of the 2017 ACM on Conference on
  Information and Knowledge Management}}. ACM, \bibinfo{pages}{2411--2414}.
\newblock


\bibitem[\protect\citeauthoryear{Yu, Liu, Wu, Wang, and Tan}{Yu
  et~al\mbox{.}}{2016}]%
        {yu2016dynamic}
\bibfield{author}{\bibinfo{person}{Feng Yu}, \bibinfo{person}{Qiang Liu},
  \bibinfo{person}{Shu Wu}, \bibinfo{person}{Liang Wang}, {and}
  \bibinfo{person}{Tieniu Tan}.} \bibinfo{year}{2016}\natexlab{}.
\newblock \showarticletitle{A dynamic recurrent model for next basket
  recommendation}. In \bibinfo{booktitle}{\emph{Proceedings of the 39th
  International ACM SIGIR conference on Research and Development in Information
  Retrieval}}. ACM, \bibinfo{pages}{729--732}.
\newblock


\bibitem[\protect\citeauthoryear{Zhang, Xiong, Su, Duan, and Zhang}{Zhang
  et~al\mbox{.}}{2016}]%
        {zhang2016variational}
\bibfield{author}{\bibinfo{person}{Biao Zhang}, \bibinfo{person}{Deyi Xiong},
  \bibinfo{person}{Jinsong Su}, \bibinfo{person}{Hong Duan}, {and}
  \bibinfo{person}{Min Zhang}.} \bibinfo{year}{2016}\natexlab{}.
\newblock \showarticletitle{Variational neural machine translation}.
\newblock \bibinfo{journal}{\emph{arXiv preprint arXiv:1605.07869}}
  (\bibinfo{year}{2016}).
\newblock


\bibitem[\protect\citeauthoryear{Zhang, Tay, Yao, Sun, and An}{Zhang
  et~al\mbox{.}}{2019}]%
        {zhang2019next}
\bibfield{author}{\bibinfo{person}{Shuai Zhang}, \bibinfo{person}{Yi Tay},
  \bibinfo{person}{Lina Yao}, \bibinfo{person}{Aixin Sun}, {and}
  \bibinfo{person}{Jake An}.} \bibinfo{year}{2019}\natexlab{}.
\newblock \showarticletitle{Next Item Recommendation with Self-Attentive Metric
  Learning}. In \bibinfo{booktitle}{\emph{Thirty-Third AAAI Conference on
  Artificial Intelligence}}, Vol.~\bibinfo{volume}{9}.
\newblock


\bibitem[\protect\citeauthoryear{Zhou, Bai, Song, Liu, Zhao, Chen, and
  Gao}{Zhou et~al\mbox{.}}{2018}]%
        {zhou2018atrank}
\bibfield{author}{\bibinfo{person}{Chang Zhou}, \bibinfo{person}{Jinze Bai},
  \bibinfo{person}{Junshuai Song}, \bibinfo{person}{Xiaofei Liu},
  \bibinfo{person}{Zhengchao Zhao}, \bibinfo{person}{Xiusi Chen}, {and}
  \bibinfo{person}{Jun Gao}.} \bibinfo{year}{2018}\natexlab{}.
\newblock \showarticletitle{ATRank: An attention-based user behavior modeling
  framework for recommendation}. In \bibinfo{booktitle}{\emph{Thirty-Second
  AAAI Conference on Artificial Intelligence}}.
\newblock


\end{thebibliography}

\end{document}